\documentclass[11pt,a4paper]{article}
\usepackage[T1]{fontenc}
\usepackage[utf8]{inputenc}
\usepackage[margin=2.5cm]{geometry}
\usepackage{amsmath,amssymb,mathrsfs,booktabs,array,xcolor}
\usepackage{microtype}
\usepackage[colorlinks=true,linkcolor=blue,citecolor=blue,urlcolor=blue]{hyperref}
\hypersetup{pdftitle={First-order origins of second-order supersymmetry},pdfauthor={Stevan Cebrian and Mikhail S. Plyushchay}}
\newcommand{\dd}{\mathrm d}
\newcommand{\id}{\mathbf 1}
\newcommand{\diag}{\operatorname{diag}}
\newcommand{\tr}{\operatorname{tr}}
\newcommand{\Bol}{\mathcal B}
\makeatletter
\newcommand{\appendixlabel}[1]{%
  \begingroup\phantomsection\def\@currentlabel{\thesection}%
  \label{#1}\endgroup}
\makeatother
\numberwithin{equation}{section}
\title{First-order origins of second-order supersymmetry}
\author{Stevan Cebrian and Mikhail S. Plyushchay\\[6pt]
{\normalsize\itshape Departamento de F\'isica, Universidad de Santiago de Chile,}\\
{\normalsize\itshape Avenida V\'ictor Jara 3493, Santiago, Chile}\\[6pt]
{\small\href{mailto:stevan.cebrian@usach.cl}{\texttt{stevan.cebrian@usach.cl}},
\href{mailto:mikhail.plyushchay@usach.cl}{\texttt{mikhail.plyushchay@usach.cl}}}}
\date{}
\begin{document}
\maketitle

\begin{abstract}
One-dimensional second-order supersymmetry is constructed from conserved
first-order operators on a surface with a cyclic coordinate. In Witten's
mechanics with two fermionic modes, a centrally extended $N=4$ algebra
permits a conserved product connecting the zero- and two-fermion sectors.
Cyclic reduction gives the second-order supercharge and the correlated
quantum corrections in the charge and partner Hamiltonians. Isothermal
coordinates identify the auxiliary Hill equation with the separated
Laplace or wave equation for the gauge-transformed charge kernel,
and express the correction as a Schwarzian derivative.
The central momentum fixes the quadratic invariant $C$: the Riemannian
construction gives $C\leq0$, and its indefinite counterpart gives
$C\geq0$, with a separately chosen positive adjoint after reduction.
Diagonalizing the one-fermion intermediate recovers real scalar
factorizations for $C\leq0$ and complex ones for $C>0$.
Pauli reductions provide a second first-order mechanism; on the
hyperbolic plane they reveal a curvature-dependent invariant and
additional charges of a regular free-particle--P\"oschl--Teller pair
in a selected momentum sector. In one dimension, a formally Hermitian
two-component intermediate with a positive inner product yields local
first-order chains for arbitrary $C$. For a constant superpotential, a real
transverse deformation on a flat cylinder also gives a first-order
surface realization of $C>0$ with a positive inner product.
\end{abstract}

\section{Introduction}
\label{sec:introduction}

The order $n$ of one-dimensional supersymmetry refers to the degree of
its supercharges in momentum, or to their differential order after
quantization \cite{AndrianovIoffeSpiridonov1993,AndrianovIoffe2012}.
For $n=1$, factorization
gives conserved quantum charges for an arbitrary real superpotential
$W(x)$. Classical nonlinear supersymmetry also allows arbitrary
superpotentials. Direct quantization, however, generally destroys
conservation of the nonlinear charges: this is the supersymmetry anomaly
\cite{Plyushchay1999,KlishevichPlyushchay2001}. In holomorphic higher-order constructions,
removing it imposes additional differential conditions on $W$.

Second-order supersymmetry admits a universal local resolution of the
quantum anomaly. For an arbitrary smooth nonvanishing $W$, correlated
corrections of order $\hbar^2$ in the Hamiltonians and supercharges
quantize the general classical quadratic family while preserving its
polynomial superalgebra. General classical and quantum constructions
also exist at third order, including the quantum solution of
Ref.~\cite{IoffeNishnianidze2004}; their existence does not by itself
provide an analogous universal correction prescription.
The second-order correction and its Schwarzian representation are
discussed in Ref.~\cite{Plyushchay2017}. Its amplitude form has the
structure of the Madelung--Bohm quantum potential
\cite{Madelung1927,Bohm1952}. The geometry of the Madelung transformation
itself, including its momentum-map interpretation, is developed in
Ref.~\cite{KhesinModin2026}. Extension through a zero of $W$ requires
additional conditions ensuring regularity of the complete operators.

Nonlinear algebras of conserved quantities also play a direct role in
bosonic mechanics. The hidden integrals of the Kepler and oscillator
problems on a sphere obey curvature-dependent polynomial relations
\cite{Higgs1979,Leemon1979}. In quantum superintegrable systems,
polynomial algebras organize degeneracies and provide algebraic methods
for determining energy levels \cite{Daskaloyannis2001}; their classical
and quantum manifestations are reviewed in
Ref.~\cite{MillerPostWinternitz2013}.

For spinning particles, Killing--Yano two-forms generate additional odd
charges whose brackets can yield hidden bosonic integrals
\cite{GibbonsRietdijkVanHolten1993,VanHolten1995}.
This geometry also underlies an additional
supercharge of the fermion--monopole system, whose full supersymmetry
algebra is nonlinear \cite{DeJongheEtAl1995,Plyushchay2000}.
At the quantum level, Killing--Yano tensors yield first-order Dirac symmetry operators
whose lower-order terms ensure exact quantum symmetry
\cite{Cariglia2004}. A broad account of classical and quantum hidden
symmetries is given in Ref.~\cite{Cariglia2014}; their role in black-hole
integrability is reviewed in Ref.~\cite{FrolovEtAl2017}.

Dimensional reduction supplies concrete bosonic examples in which
nonlinear algebras arise from Lie-algebra generators. In the reduction
of the four-dimensional isotropic oscillator to a two-dimensional
singular oscillator, invariant composites of the parent $\mathfrak u(4)$
generators descend to conserved operators realizing the polynomial Higgs--Hahn algebra
\cite{FrappatEtAl2019}. Similarly, composites of first-order
$\mathfrak o(6)$ rotation generators give the Racah symmetry algebra
of a reduced superintegrable model on $S^2$
\cite{GaboriaudEtAl2018}. A related geometrization connects the Higgs
oscillator to a Klein--Gordon equation in anti-de Sitter spacetime:
AdS isometries give spectrum-generating operators whose quadratic
combinations yield hidden integrals \cite{EvninNivesvivat2016}.
A further nonlocal construction yields the undeformed hidden
$\mathfrak{su}(d)$ algebra \cite{EvninNivesvivat2017}.
These examples motivate tracking conserved composites and their
algebra through reduction.

For second-order supersymmetry, the previous paper~\cite{Paper1} developed the associated projective
geometry using an auxiliary Hill equation and the related third-order
Bol (Gel'fand--Dikii) equation. Its invariant $C$ appears in
the same quadratic polynomial in the classical and quantum superalgebras.
The cases $C<0$, $C=0$ and $C>0$ correspond to two real roots, a repeated
root and no real roots, and to hyperbolic, parabolic and elliptic
projective flows. That paper also related the scalar quantum correction
to the radial measure in a Riemannian reduction. Here we construct the
complete first-order surface supercharges before reduction and derive
their algebra and conserved quadratic composite. In Witten's mechanics
with two fermionic modes \cite{Witten1982}, this composite connects the
zero- and two-fermion sectors through the two-component one-fermion
sector. Cyclic reduction gives the one-dimensional system, including
the correlated corrections in its Hamiltonians and charges. In the
Riemannian model, the four real first-order supercharges form a centrally
extended $N=4$ algebra whose cyclic momentum fixes $C\leq0$.
The indefinite construction gives the $C\geq0$ expressions; we track
the change from its signed adjoint to the positive reduced adjoint.
Isothermal coordinates make the auxiliary Hill equation a separated
surface Laplace or wave equation within the composite-charge kernel.
This connects the projective and surface descriptions without assuming
an additional symmetry of the full physical Hamiltonian.

We also relate the surface intermediate to the scalar factorizations of
Ref.~\cite{Plyushchay2017}, including their non-Hermitian intermediate for
$C>0$. A planar Pauli model provides a second mechanism: a first-order
supercharge is composed with a first-order spatial symmetry. Its
zero-field limit connects directly with the earlier construction of
nonlinear superconformal mechanics by reduction
\cite{AnabalonPlyushchay2003}. On the hyperbolic plane, we examine both
the curvature dependence of the reduced invariant and a regular
free-particle--P\"oschl--Teller pair. The latter distinguishes symmetries
inherited from the full surface from additional charges conserved only
in the selected momentum sector. More general nonlinear Pauli models allow
positive $C$, but their supercharges are already second order before
reduction. This distinction is essential to the first-order-origin
problem. Finally, we examine a positive matrix factorization for arbitrary
$C$ and a global realization with constant $W$ and $C>0$ on a flat cylinder.

Section~\ref{sec:target} specifies the one-dimensional system.
Sections~\ref{sec:witten} and \ref{sec:signature} give the Witten
construction and its two signatures; Section~\ref{sec:pauli} treats Pauli
reductions. Section~\ref{sec:limits} relates these results to scalar and
matrix factorizations. 
Section~\ref{sec:conclusions} is devoted to conclusions and outlook.
Supporting calculations are collected in the
appendices. We use unit mass and $\hbar>0$; primes on functions of $x$
mean $\dd/\dd x$. A superscript $t$ denotes transposition, without
complex conjugation; for differential operators it also reverses
derivatives and the order of factors. Continuous flow parameters are
denoted by $\tau$. Unless domains are specified, operator relations are
identities of differential expressions on smooth test functions supported
inside a regular coordinate region, with the prescribed periodic or
spin boundary conditions in angular variables. In a positive Hilbert space, a physical realization requires
self-adjoint Hamiltonians,
closed mutually adjoint charges, and compatible domains for their
products. An invariant core, when used, is a subspace preserved by the
operators and dense in the relevant graph norms; ordinary Hilbert-space
density alone does not suffice. Spectral conclusions below assume the
corresponding operator or closed quadratic-form identities.
The notation $N$ for the number of real supercharges is separate from
the order $n$ and from fermion occupation numbers.

\section{Second-order supersymmetry in one dimension}
\label{sec:target}

We recall one-dimensional second-order ($n=2$) supersymmetry, beginning
with the anomaly in direct quantization. We then give the correction
that restores charge conservation and the quadratic superalgebra for
an arbitrary smooth nonvanishing $W$. These are the operators that
the surface constructions must reproduce.

\subsection{The classical system and the quantization problem}

Let $x,p$ be a real canonical pair and $\theta^\pm$ a conjugate Grassmann
pair, with graded Poisson brackets
\begin{equation}
 \{x,p\}=1,\qquad \{\theta^+,\theta^-\}=-i,\qquad
 N=\theta^+\theta^-.
 \end{equation}
For real $W(x)$ and real constants $a,C$, take
\begin{align}
 H_{\rm cl}&=a+\frac12\left(p^2+W^2-\frac C{W^2}\right)+2W'N,
 \label{eq:classical-target}\\
 Q^{\rm cl}_+&=B\theta^+,\qquad
 B=\frac12\left[(W+ip)^2+\frac C{W^2}\right],\qquad
 Q^{\rm cl}_-=(Q^{\rm cl}_+)^*.
\end{align}
For $C\ne0$ these expressions are initially defined where $W\ne0$;
for $C=0$ the inverse powers are absent. They obey
\begin{equation}
 \{H_{\rm cl},Q^{\rm cl}_\pm\}=0,\qquad
 \{Q^{\rm cl}_+,Q^{\rm cl}_-\}=-i[(H_{\rm cl}-a)^2+C]
 \label{eq:classical-algebra}
\end{equation}
without a differential restriction on $W$. The constant $a$ fixes the
energy origin. The central parameter $C$ has dimensions of energy
squared and has vanishing Poisson brackets with all generators.

In the quantum theory $p=-i\hbar\partial_x$, and the fermionic pair gives
two partner sectors. Introduce a real scalar function $U(x)$ and write
\begin{align}
 H_\pm&=a+\frac12[-\hbar^2\partial_x^2+W^2+U\pm2\hbar W'],
 \label{eq:H}\\
 q_+&=\frac12[(\hbar\partial_x+W)^2-U],\qquad q_-=q_+^\dagger.
 \label{eq:q}
\end{align}
The adjoint is initially formal in $L^2(\dd x)$. The square in $q_+$
is an operator product and includes $\hbar W'$.
The scalar term appears as $+U/2$ in each Hamiltonian and $-U/2$
in each charge. This relative sign follows from the derivative terms
of the intertwining equation; Appendix~\ref{app:intertwiner} derives it
starting with a general second-order charge.

We assign $q_+$ to map the minus sector into the plus sector.
This assignment specifies its matrix position, but does not yet make it
a symmetry. Preservation of energy requires the intertwining identity
$H_+q_+=q_+H_-$. Direct expansion gives
\begin{equation}
 H_+q_+-q_+H_-=\frac\hbar4(\hbar^2W'''-4UW'-2U'W).
 \label{eq:defect}
\end{equation}
The direct classical choice $U=-C/W^2$ cancels the last two terms,
but leaves $\hbar^3W'''/4$. This is the quantum anomaly: for a generic
$W$, the charge is no longer conserved. The special case $W'''=0$
has no such remainder.

\subsection{The universal correction and the quantum algebra}

With $\mathcal U=-2U/\hbar^2$, the vanishing of Eq.~\eqref{eq:defect}
is equivalent to
\begin{equation}
 \Bol[W]=0,\qquad
 \Bol=\partial_x^3+2\mathcal U\partial_x+\mathcal U',\qquad
 \Bol[W]=W'''+2\mathcal U W'+\mathcal U'W.
 \label{eq:Bol}
\end{equation}
This is the third Bol operator, also called the Gel'fand--Dikii operator
in this normalization \cite{GelfandDikii1975}. For a prescribed
$\mathcal U$, it gives a linear third-order equation for $W$.
Here $W$ is prescribed, and the same equation determines $U$ up to
the integration constant $C$.
Multiplication by $2W$ produces an exact derivative:
\begin{equation}
 I[W]=2WW''-W'^2+2\mathcal U W^2=\frac{4C}{\hbar^2},
 \qquad I'[W]=2W\Bol[W].
 \label{eq:I}
\end{equation}
We choose the integration constant to be the same $C$ as in the
classical algebra. On an interval where $W$ has fixed nonzero sign,
\begin{equation}
 U=\Delta_W-\frac C{W^2},\qquad
 \Delta_W=\hbar^2\left(\frac{W''}{2W}-\frac{W'^2}{4W^2}\right).
 \label{eq:correction}
\end{equation}
Thus an arbitrary smooth nonvanishing $W$ has an anomaly-free
quantization. The purely quantum term enters $H_\pm$ as $+\Delta_W/2$
and $q_\pm$ as $-\Delta_W/2$. Besides conservation, it yields
\begin{equation}
 H_+q_+=q_+H_-,\qquad
 q_-q_+=(H_--a)^2+C,\qquad q_+q_-=(H_+-a)^2+C.
 \label{eq:poly}
\end{equation}
Once this compatibility condition is satisfied, $q_+$ sends a
minus-sector energy solution to a plus-sector solution at the same
energy, or annihilates it. In the initial ansatz, $W$ and $U$ therefore
cannot be chosen independently.

Charge conservation alone does not ensure that quantization leaves the
polynomial parameter $C$ unchanged. For example, $W=\omega x$ and
$U=-C_0/W^2$ give zero
defect because $W'''=0$, but Eq.~\eqref{eq:I} fixes the quantum
closure constant to
\begin{equation}
 C=C_0-\frac{\hbar^2\omega^2}{4}.
\end{equation}
The prescription~\eqref{eq:correction}, with the correlated correction
in both Hamiltonians and charges, instead keeps the specified $C$
unchanged.

The correction also has the Madelung--Bohm amplitude form. On a
fixed-sign interval put $R_W=\sqrt{|W|}$. Then
\begin{equation}
 \Delta_W=\hbar^2\frac{R_W''}{R_W},\qquad
 \frac{\Delta_W}{2}=-Q[R_W],\qquad
 Q[R]=-\frac{\hbar^2}{2}\frac{R''}{R}.
 \label{eq:bohm-main}
\end{equation}
The contribution to the potential is minus the mass-one quantum
potential evaluated on $R_W$. For physical wave amplitudes, the quantum
potential also encodes local wave geometry, as illustrated by its
relation to superoscillations \cite{Berry2020}. Here, $R_W$ has a
precise auxiliary meaning: substituting $W=sR_W^2$, with fixed $s=\pm1$,
into the first integral~\eqref{eq:I} gives the Ermakov--Pinney
equation~\cite{Pinney1950}
\begin{equation}
 R_W''+\frac{\mathcal U}{2}R_W
 =\frac{C}{\hbar^2R_W^3}.
 \label{eq:ermakov-pinney}
\end{equation}
It is associated with the auxiliary Hill equation
$\psi''+\mathcal U\psi/2=0$, discussed below in
Eq.~\eqref{eq:hill-main}. For $C=0$, $R_W$ itself solves this linear
equation. Thus $W=\pm\psi^2$ for a real Hill solution gives
$R_W=|\psi|$ and selects $C=0$.
For $C>0$, $R_W$ is instead the modulus of a complex Hill solution:
\[
 \psi_{\rm c}=R_W e^{i\vartheta},\qquad
 \vartheta'=\frac{\sqrt C}{\hbar R_W^2},\qquad
 W=\pm|\psi_{\rm c}|^2.
\]
Indeed, the imaginary part of the Hill equation gives
$(R_W^2\vartheta')'=0$, while its real part gives
Eq.~\eqref{eq:ermakov-pinney}. For $C<0$, one can locally factor
$W=uv$ into two independent real Hill solutions, so
$R_W=\sqrt{|uv|}$. This generalized amplitude cannot be the modulus
of a complex solution of the same real Hill equation: the coefficient
of $R_W^{-3}$ obtained from such a modulus is the nonnegative square
$(R_W^2\vartheta')^2$.
The amplitude in Eq.~\eqref{eq:bohm-main} therefore belongs to the
auxiliary Hill--Ermakov--Pinney system and is fixed by $W$; it is not
the amplitude of an arbitrary physical state of $H_\pm$.

The matrix Hamiltonian, nilpotent charges and grading are
\begin{equation}
 \mathbb H=\begin{pmatrix}H_+&0\\0&H_-\end{pmatrix},\quad
 \mathbb Q_+=\begin{pmatrix}0&q_+\\0&0\end{pmatrix},\quad
 \mathbb Q_-=\mathbb Q_+^\dagger,\quad
 \Gamma=\diag(1,-1).
 \label{eq:matrix-target}
\end{equation}
The two components are the plus and minus partner sectors. Their superalgebra is
\begin{equation}
 \mathbb Q_\pm^2=0,\qquad [\mathbb H,\mathbb Q_\pm]=0,\qquad
 \{\mathbb Q_+,\mathbb Q_-\}=(\mathbb H-a)^2+C.
 \label{eq:superalgebra}
\end{equation}
Here $C\id$ commutes with all quantum generators. With the fermionic
normalization absorbed into the matrix charges, the polynomial is
exactly the classical polynomial in Eq.~\eqref{eq:classical-algebra}.
The operators acquire quantum corrections, while this algebraic
structure is preserved.

\subsection{Three regimes of \texorpdfstring{$C$}{C}: roots and physical states}

For mutually adjoint charges and a normalized energy eigenstate in their
common domain, Eq.~\eqref{eq:superalgebra} implies
\begin{equation}
 \|\mathbb Q_+\Psi_E\|^2+\|\mathbb Q_-\Psi_E\|^2
 =P(E),\qquad P(E)=(E-a)^2+C\geq0.
 \label{eq:positivity}
\end{equation}
A supersymmetry singlet is annihilated by both charges, so its energy
must be a zero of $P$. One charge alone also annihilates an entire
grading sector, including members of ordinary doublets.
For $C<0$ define $\beta=\sqrt{-C}>0$. The three possibilities are
\begin{equation}
 \begin{array}{c|c|c}
 &P(E)&\text{candidate singlet energies}\\ \hline
 C=-\beta^2<0&(E-a-\beta)(E-a+\beta)&a-\beta,\ a+\beta\\
 C=0&(E-a)^2&a\\
 C>0&(E-a)^2+C&\text{none on the real axis}
 \end{array}.
 \label{eq:signs}
\end{equation}
For a self-adjoint realization of the polynomial superalgebra,
positivity excludes the whole spectral interval $(a-\beta,a+\beta)$
when $C<0$, but does not select between the two remaining energy
regions. This statement includes continuous spectrum; its operator
assumptions are explained in Appendix~\ref{app:quantum-domains}.
For $C>0$, compatible closure and intertwining imply a unitary,
generally nonlocal equivalence of the partner Hamiltonians through the
polar part of $q_+$, also described there. In every case the algebra
specifies candidate singlets; normalizability and the operator domains
determine whether they are actual states.

The distinction can already be seen at the classical level in the
ordinary complex coefficient $B$ of the Grassmann charge. Removing
the nilpotent fermionic term, define
$h=H_{\rm cl,body}-a=\frac12(p^2+W^2-C/W^2)$; then
$|B|^2=h^2+C$. For $C=-\beta^2<0$,
\begin{equation}
 h=\beta+\frac12\left[p^2+\left(W-\frac\beta W\right)^2\right]
 \geq\beta.
 \label{eq:classical-bound}
\end{equation}
Only the upper root is classically accessible, and only where
$p=0$ and $W^2=\beta$. Moreover,
$2B=(W-\beta/W+ip)(W+\beta/W+ip)$. The second factor never vanishes
for real $p,W\ne0$. A local canonical transformation absorbs its phase
into the fermions, with compensating nilpotent shifts of $x,p$.
Dividing the transformed charge by $\sqrt{H_{\rm cl}-a+\beta}$, whose
bosonic part is strictly positive, then gives a charge linear in the
transformed momentum. The Hamiltonian takes the ordinary $n=1$ form;
the steps are given in Appendix~\ref{app:classical}.

For $C=0$, $B=(W+ip)^2/2$ is holomorphic \cite{KlishevichPlyushchay2001}. Near a simple zero of $W$
at $p=0$, it has a double phase-space zero; such a point need not lie
in every chosen interval. For $C>0$, $B$ has no zero, and its phase
and modulus can be removed locally by a canonical transformation and
a regular charge normalization. The resulting conserved fermions give
the $n=0$ normal form. These statements classify charges on regular
patches; they do not remove singularities of the original bosonic
potential. Appendix~\ref{app:classical} gives the transformations and
explains why they need not extend to quantum equivalences.

Quantum mechanics can realize both roots when $C<0$. Take
$W=\omega x$, $\omega>0$, and $C=-\hbar^2\omega^2/4$, so that
Eq.~\eqref{eq:correction} gives $U=0$. On the full line,
\begin{equation}
 H_-=a+\hbar\omega(N-\tfrac12),\qquad
 H_+=a+\hbar\omega(N+\tfrac32),\qquad
 q_+=\hbar\omega(b^-)^2,
 \label{eq:oscillator}
\end{equation}
where $b^-=(\hbar\partial_x+\omega x)/\sqrt{2\hbar\omega}$,
$b^+=(b^-)^\dagger$, and now $N=b^+b^-$ is the oscillator number
operator. 
The oscillator states \(|0\rangle\) and \(|1\rangle\) belong to \(\ker q_+\).
Consequently, the corresponding lower-sector states
\(\Psi_n=(0,|n\rangle)^t\), \(n=0,1\), are annihilated by both matrix
supercharges \(\mathbb Q_+\) and \(\mathbb Q_-\). They are physical
supersymmetry singlets with energies \(a-\hbar\omega/2\) and
\(a+\hbar\omega/2\), respectively, although only the upper root is accessible
in the corresponding classical realization.

At $x=0$, the separate terms $\Delta_W=-\hbar^2/(4x^2)$ and
$-C/W^2=\hbar^2/(4x^2)$ diverge, but their sum $U$ vanishes.
Thus a singular divided formula need not define a singular quantum
operator. More generally, regular continuation through a simple zero
requires $C=-\hbar^2W'(x_0)^2/4$; a double zero requires $C=0$ and
$W'''(x_0)=0$. A nontrivial $W$ cannot have a triple or higher zero
when $U$ is smooth, by uniqueness of Eq.~\eqref{eq:Bol}. On a connected
regular interval, any zeros are isolated and their type is fixed by
$C$: simple for $C<0$, double for $C=0$, and absent for $C>0$.
These conditions concern continuation through zeros, rather than SUSY
on punctured intervals. Appendix~\ref{app:zeros} gives the argument
and the finite limiting values of $U$.

At an admissible simple zero, the same correction that preserves the
quadratic superalgebra cancels the inverse-power singularities in the
Hamiltonians and supercharges; the classical potential has a repulsive
barrier there. For an $\hbar$-independent $W$, regularity selects
$C\propto\hbar^2$, so preserving this cancellation as $\hbar\to0$
requires $C\to0$. At an admissible double zero, $C=0$ and the
classical inverse-power term is already absent. In both cases, smooth
coefficients permit local continuation, while a physical realization
still requires compatible operator domains.

\subsection{The auxiliary projective equation and its physical role}

The second-order equation underlying the Bol operator is
\begin{equation}
 \mathscr L=\partial_x^2+\frac12\mathcal U,\qquad \mathscr L\psi=0.
 \label{eq:hill-main}
\end{equation}
It is an auxiliary Hill equation, rather than a physical energy
eigenvalue equation. Its relation to the supercharges follows by
choosing $\mathcal S'(x)=W(x)$:
\begin{equation}
 q_+=\frac{\hbar^2}{2}e^{-\mathcal S/\hbar}\mathscr L
                  e^{\mathcal S/\hbar},\qquad
 q_-=\frac{\hbar^2}{2}e^{\mathcal S/\hbar}\mathscr L
                  e^{-\mathcal S/\hbar}.
 \label{eq:hill-charge-main}
\end{equation}
Every Hill solution therefore gives local functions
$e^{-\mathcal S/\hbar}\psi\in\ker q_+$ and
$e^{\mathcal S/\hbar}\psi\in\ker q_-$. Whether these functions are
square-integrable and satisfy the chosen boundary conditions remains
a separate question: these kernels initially denote local solution spaces.
Such finite-dimensional invariant spaces also underlie the relation
between higher-order supersymmetry and quasi-solvability
\cite{AoyamaSatoTanaka2001}. In the terminology of that reference,
$\mathcal N$-fold supersymmetry counts the order of the charge, not
the number $N$ of real supercharges used here.

Under $x=\phi(y)$ with $\phi'>0$, a Hill solution transforms as
$\widetilde\psi(y)=\phi'(y)^{-1/2}\psi(\phi(y))$, a density of weight
$-1/2$. Its transformed equation has coefficient
\begin{equation}
 \widetilde{\mathcal U}(y)=\phi'(y)^2\mathcal U(\phi(y))+\{\phi;y\},
 \qquad \{\phi;y\}=\frac{\phi'''}{\phi'}
                   -\frac32\left(\frac{\phi''}{\phi'}\right)^2.
 \label{eq:projective-law-main}
\end{equation}
Primes on $\phi$ mean $y$ derivatives. The Schwarzian term makes
$\mathcal U$ a projective connection; this geometric interpretation
of Sturm--Liouville operators is discussed in
Ref.~\cite{OvsienkoTabachnikov2005}. Infinitesimally, for
$\phi(y)=y+\tau W(y)+O(\tau^2)$, comparison at the same argument gives
$\widetilde{\mathcal U}(y)-\mathcal U(y)=\tau\Bol[W](y)+O(\tau^2)$.
Hence $\Bol[W]=0$ means that the flow generated by $W\partial_x$
stabilizes this projective connection.

The corresponding Lie derivative of a weight-$-1/2$ density is
\begin{equation}
 \mathfrak D_W\psi=W\partial_x\psi-\tfrac12W'\psi.
 \label{eq:lie-main}
\end{equation}
The negative half-weight fixes the minus sign in its second term.
When $\Bol[W]=0$, this operator maps Hill solutions to Hill solutions.
Choose real independent solutions $\psi_1,\psi_2$ with constant
nonzero Wronskian $\omega_0=\psi_1\psi_2'-\psi_1'\psi_2$.
Their three products $\psi_1^2,\psi_1\psi_2,\psi_2^2$ span the
solutions of the third-order equation. Thus
\begin{equation}
 W=\boldsymbol\psi^tM\boldsymbol\psi,\qquad
 \boldsymbol\psi=\begin{pmatrix}\psi_1\\\psi_2\end{pmatrix},\qquad
 M=\begin{pmatrix}m_{11}&m_{12}\\m_{12}&m_{22}\end{pmatrix},
 \label{eq:projective-data}
\end{equation}
where $M$ is a constant real symmetric matrix.
For a Hill solution $\xi=c_1\psi_1+c_2\psi_2$, let
$c=(c_1,c_2)^t$ be its constant coefficient column. Then
\begin{equation}
 \mathfrak D_W(\boldsymbol\psi^t c)=\boldsymbol\psi^tAc,\qquad
 A=\omega_0MJ,\qquad J=\begin{pmatrix}0&1\\-1&0\end{pmatrix},
 \qquad A^2=-\frac C{\hbar^2}\id.
 \label{eq:A}
\end{equation}
The identity for $A^2$ is the restriction to Hill solutions of the
differential identity
\begin{equation}
 \mathfrak D_W^{\,2}-W^2\mathscr L
 =-\frac14 I[W]\,\id=-\frac{C}{\hbar^2}\id.
 \label{eq:lie-hill-identity}
\end{equation}
It follows from the first integral~\eqref{eq:I} and holds on arbitrary
smooth functions; the term $W^2\mathscr L$ vanishes on the Hill kernel.
Together with the conjugations~\eqref{eq:hill-charge-main} and the
first-order gauge identities in Section~\ref{sec:ordering}, it connects
the Lie derivative with the Hill operator, the intertwiners and the
associated first-order operators.

For two solutions with coefficient columns $c,d$, their conserved
Wronskian is $\omega_0c^tJd$. The identity $A^tJ+JA=0$ says that
the induced flow $e^{\tau A}$ preserves this symplectic pairing.
Its action on the ratio of coefficients is a M\"obius transformation,
whose type is fixed by $C$.

The relation to the quadratic $\mathfrak{sl}(2,\mathbb R)$ Casimir
requires distinguishing the representation from the selected generator.
On a patch with $\psi_2\ne0$, use the projective coordinate
$f=\psi_1/\psi_2$, choosing the sign of $\psi_1$ so that
$f'=-\omega_0/\psi_2^2>0$. The transformed Hill equation is
$\partial_f^2\widetilde\psi=0$, with solution space spanned by $1,f$.
Its stabilizer algebra is spanned by
$\partial_f,f\partial_f,f^2\partial_f$.
On a density of weight $\rho$, the Lie derivative along
$v(f)\partial_f$ is
$\mathfrak D_v^{(\rho)}=v\partial_f+\rho\,(\partial_f v)$.
The corresponding generators are therefore
\begin{equation}
 J_-^{(\rho)}=\partial_f,\qquad
 J_0^{(\rho)}=f\partial_f+\rho,\qquad
 J_+^{(\rho)}=f^2\partial_f+2\rho f.
 \label{eq:density-sl2}
\end{equation}
With $[J_0,J_\pm]=\pm J_\pm$ and $[J_-,J_+]=2J_0$,
these real differential generators have Casimir
\begin{equation}
 \mathcal K^{(\rho)}
 =\bigl(J_0^{(\rho)}\bigr)^2
 -\frac12\bigl\{J_+^{(\rho)},J_-^{(\rho)}\bigr\}
 =\rho(\rho-1)\id.
 \label{eq:density-casimir}
\end{equation}
The Hill solution space has weight $-1/2$ and Casimir $3/4$;
its quadratic products form the three-dimensional Bol
(Gel'fand--Dikii) solution space, whose vector-field coefficients
have weight $-1$ and Casimir $2$.
By contrast, $C$ is the basis-independent quadratic invariant of the generator
selected by $W$ in the Hill representation:
\begin{equation}
 C=\hbar^2\omega_0^2\det M
  =\hbar^2\det A=-\frac{\hbar^2}{2}\tr(A^2).
 \label{eq:selected-generator-invariant}
\end{equation}
Changing $W$ at fixed $\mathcal U$ can change $C$ while leaving
the representation Casimir unchanged.
For example, at fixed $\mathcal U=0$, let $L>0$ be a length and
$w_*>0$ a momentum scale. The Bol solutions
$W=w_*$, $w_*x/L$ and $w_*[1+(x/L)^2]$ give, respectively,
$C=0$, $-(\hbar w_*/2L)^2$ and $(\hbar w_*/L)^2$, while the
Hill representation has Casimir $3/4$ in every case.

The three-dimensional quadratic data also carry a Lorentzian structure.
Writing $m_{11}=\xi_0+\xi_1$, $m_{22}=\xi_0-\xi_1$ and
$m_{12}=\xi_2$ gives $\det M=\xi_0^2-\xi_1^2-\xi_2^2$.
The congruence action $M\mapsto GMG^t$, $G\in SL(2,\mathbb R)$,
preserves this form and realizes the double cover of $SO^+(2,1)$.
Thus $C<0$, $C>0$ and $C=0$ with $W\not\equiv0$ correspond,
respectively, to spacelike, timelike and null data, and to hyperbolic,
elliptic and parabolic flows. An elliptic flow is a rotation within
this Lorentz group; it does not replace that group by $SO(3)$.
This structure belongs to the auxiliary solution space, not to the
particle's physical spacetime.

The same auxiliary matrix determines the Hamiltonian restricted to
each formal charge kernel. In the bases
$e^{-\mathcal S/\hbar}\psi_{1,2}$ and
$e^{\mathcal S/\hbar}\psi_{1,2}$ of the local charge kernels,
$H_-$ and $H_+$ are represented by $a\id+\hbar A$ and
$a\id-\hbar A$, respectively. Their formal eigenvalues give the
candidate singlet energies in Eq.~\eqref{eq:signs}. This is a restriction
to two-dimensional solution spaces, not a reduction of differential
order on arbitrary wavefunctions. The auxiliary flow parameter is
likewise not the particle's physical time.
Appendix~\ref{app:projective} derives the matrix representation and
the kernel restriction, including their domain qualifications.

\section{A first-order surface origin and the quantum correction}
\label{sec:witten}

The construction below uses a particle moving on a two-dimensional surface,
with two fermionic modes. Its zero-fermion sector will give the minus
partner $H_-$, and its two-fermion sector the plus partner $H_+$.
Two successive first-order supersymmetry transformations connect these
sectors through the two-component one-fermion sector. Their product
becomes $q_+$ after the angular momentum is fixed. This explains both
the second differential order and the choice of the two partner sectors.

\subsection{The surface Hamiltonian and its fermionic components}

Let $x$ be a distance coordinate on an open interval $I$ and
$\varphi$ an angle of period $2\pi$. Take
\begin{equation}
 \dd s^2=\dd x^2+F(x)^2\dd\varphi^2,\qquad F(x)>0.
 \label{eq:metric}
\end{equation}
The orbit at fixed $x$ has circumference $2\pi F(x)$; thus $F(x)$
is its circumferential radius, not in general a proper radial distance.
This coordinate region has topology $I\times S^1$, that of an open
cylinder. A global extension through an endpoint where $F$ vanishes
and the circular orbit collapses may have a different topology.
Primes in this section denote $\dd/\dd x$. Two fermionic modes,
$\{f_i,f_j^\dagger\}=\delta_{ij}$, give four occupation states with
amplitudes $(\Psi_0,\Psi_x,\Psi_\varphi,\Psi_2)^t$:
\begin{equation}
 \Psi_0|0\rangle+\Psi_x f_x^\dagger|0\rangle
 +\Psi_\varphi f_\varphi^\dagger|0\rangle
 +\Psi_2 f_x^\dagger f_\varphi^\dagger|0\rangle.
\end{equation}
Each amplitude is a function of $x,\varphi$. Equivalently, the occupation
states can be represented by the forms $1,e^x,e^\varphi,e^x\wedge e^\varphi$,
where $e^x=\dd x$ and $e^\varphi=F\dd\varphi$ are orthonormal. The norm is
\begin{equation}
 \|\Psi\|^2=\int F\dd x\dd\varphi\,
 (|\Psi_0|^2+|\Psi_x|^2+|\Psi_\varphi|^2+|\Psi_2|^2).
\end{equation}
Form degree is fermion occupation number. The fermion parity is
$(-1)^{N_f}=\diag(1,-1,-1,1)$ in this component order.

Choose a real $\chi(x)$ with $\chi'=W$, where $W$ is the superpotential
of Section~\ref{sec:target}. The function $\chi$ has dimensions of action
and $W$ of momentum. Witten's first-order differential and Hamiltonian are
\begin{equation}
 d_\chi=\hbar\dd+\dd\chi\wedge
       =\hbar e^{-\chi/\hbar}\dd e^{\chi/\hbar},\qquad
 Q=-i d_\chi,\qquad
 \mathcal H_W=\tfrac12\{Q,Q^\dagger\}.
 \label{eq:witten-basic}
\end{equation}
Here $\dd$ is the exterior derivative on the two-dimensional surface. 
For a scalar function $f(x,\varphi)$,
$\dd f=\partial_x f\,\dd x+\partial_\varphi f\,\dd\varphi$,
while $\dd\chi\wedge$ creates
one fermion by multiplying a form by $W e^x$ with the wedge product.
The adjoint uses the positive norm above. Since $\dd^2=0$,
$d_\chi^2=0$ and $[\mathcal H_W,Q]=0$. The Hamiltonian preserves
fermion occupation number: its blocks have dimensions $1,2,1$.
For example,
\begin{equation}
 \mathcal H_W^{(0)}
 =-\frac{\hbar^2}{2}\left[\frac1F\partial_x(F\partial_x)
                 +\frac1{F^2}\partial_\varphi^2\right]
   +\frac{W^2}{2}-\frac\hbar2\left(W'+\frac{F'}F W\right).
\end{equation}
This displays the surface kinetic energy, scalar potential $W^2/2$,
and supersymmetric term. In $\mathcal H_W^{(2)}$ the last term has
the opposite sign. So far $W$ and $F$ are independent.

\subsection{The second first-order charge and the centrally extended algebra}

To specify the action on the four components, define
\begin{equation}
 X=\hbar\partial_x+W,\qquad Y=\frac\hbar F\partial_\varphi,\qquad
 T=\hbar\partial_x+W+\hbar\frac{F'}F.
 \label{eq:XYT}
\end{equation}
The scalar component $\Psi_0$ is mapped into the one-fermion components,
and these into the two-fermion component, as follows:
\begin{equation}
 \Psi_0\ \xrightarrow{\,d_\chi\,}\
 \begin{pmatrix}X\Psi_0\\Y\Psi_0\end{pmatrix},\qquad
 \begin{pmatrix}\Psi_x\\\Psi_\varphi\end{pmatrix}
 \ \xrightarrow{\,d_\chi\,}\ -Y\Psi_x+T\Psi_\varphi.
\end{equation}
There is no further raising action on $\Psi_2$. The term $F'/F$ in $T$
is fixed by $\dd e^\varphi=(F'/F)e^x\wedge e^\varphi$; it comes from
differentiating the local basis of the one-fermion states.

A second application of $d_\chi$ vanishes. A different first-order step
is supplied by the rotation
$\mathcal R(\Psi_x,\Psi_\varphi)^t=(-\Psi_\varphi,\Psi_x)^t$.
Let $\mathcal R$ act as the identity on $\Psi_0,\Psi_2$ and define
$d_\chi^c=\mathcal R d_\chi\mathcal R^{-1}$. Its blocks are
\begin{equation}
 \begin{array}{c|cc}
 &0\longrightarrow1&1\longrightarrow2\\ \hline
 d_\chi&(X,Y)^t&(-Y,T)\\
 d_\chi^c&(-Y,X)^t&(-T,-Y)
 \end{array}.
 \label{eq:raising-blocks}
\end{equation}
For example, $d_\chi^c$ sends $\Psi_0$ to
$(-Y\Psi_0,X\Psi_0)^t$. The identity $TY=YX$ ensures its nilpotency. The two products
have opposite $0\to2$ coefficients, $TX+Y^2$ and $-TX-Y^2$,
so $\{d_\chi,d_\chi^c\}=0$.

The rotated differential has its own Hamiltonian
$\mathcal H_c=\{d_\chi^c,(d_\chi^c)^\dagger\}/2
=\mathcal R\mathcal H_W\mathcal R^{-1}$.
For both first-order charges to belong to the linear supersymmetry
used here, we impose $\mathcal H_c=\mathcal H_W$.
Their zero- and two-fermion blocks already agree. In the one-fermion block,
\begin{equation}
 \mathcal H_c^{(1)}-\mathcal H_W^{(1)}
 =\hbar\diag(-\mathfrak m,\mathfrak m),\qquad \mathfrak m=W'-\frac{F'}F W.
 \label{eq:middle-mismatch}
\end{equation}
Thus equality of the two full Hamiltonians is equivalent to
\begin{equation}
 \mathfrak m=0\quad\Longleftrightarrow\quad
 \left(\frac WF\right)'=0\quad\Longleftrightarrow\quad W=\Omega F,
 \label{eq:WF}
\end{equation}
where $\Omega\ne0$ is a real constant. The condition fixes the interaction
on a chosen surface. Conversely, any smooth nonvanishing $W$ is realized
locally on a fixed-sign interval by choosing $F=W/\Omega>0$.
Global existence of $\chi$, geodesic completeness, smooth completion
at an axis, and endpoint domains require separate analysis.
The global deformation on a torus is described in
Appendix~\ref{app:surface-global}.

The covariant Hessian $\nabla_i\nabla_j\chi$ has orthonormal components
$W'$ and $(F'/F)W$: the angular component follows from
$\Gamma^x_{\varphi\varphi}=-FF'$, even though
$\partial_\varphi\chi=0$. Thus Eq.~\eqref{eq:WF} is equivalent to
$\nabla_i\nabla_j\chi=W'g_{ij}$. The gradient
$Z_\chi=\nabla\chi=W\partial_x$ is consequently conformal Killing,
\begin{equation}
 \mathcal L_{Z_\chi}g_{ij}=2W'g_{ij},
 \qquad \mathcal JZ_\chi=\frac WF\partial_\varphi
                  =\Omega\partial_\varphi,
 \label{eq:conformal-gradient}
\end{equation}
where $\mathcal L_{Z_\chi}$ is the Lie derivative and $\mathcal J$ rotates
tangent vectors by $+\pi/2$ in the oriented orthonormal frame.
The rotated gradient is Killing; its quantum generator $\Omega K$
is the central operator below. The gradient $Z_\chi$ itself need not be
Killing.

With $\widetilde Q=d_\chi^c$ and $K=-i\hbar\partial_\varphi$, direct
composition gives
\begin{align}
 Q^2&=\widetilde Q^2=\{Q,\widetilde Q\}=0,\nonumber\\
 \{Q,Q^\dagger\}&=\{\widetilde Q,\widetilde Q^\dagger\}=2\mathcal H_W,
 \qquad
 \{Q,\widetilde Q^\dagger\}=\{\widetilde Q,Q^\dagger\}=2\Omega K.
 \label{eq:witten-algebra}
\end{align}
The phase factor \(-i\) in the definition \(Q=-i\,d_\chi\) is chosen so that both mixed anticommutators in the preceding equation equal the Hermitian operator \(2\Omega K\).
All charges commute with $\mathcal H_W$ and $K$.
This is centrally extended $N=4$ supersymmetric quantum mechanics with
first-order ($n=1$) supercharges. To display the four Hermitian generators, set
\begin{equation}
 S_\sigma=\frac{Q+\sigma\widetilde Q}{\sqrt2},\qquad
 Q_{\sigma1}=\frac{S_\sigma+S_\sigma^\dagger}{\sqrt2},\qquad
 Q_{\sigma2}=\frac{S_\sigma-S_\sigma^\dagger}{i\sqrt2},
 \qquad \sigma=\pm1.
\end{equation}
The four Hermitian operators satisfy
\begin{equation}
 \{Q_{\sigma a},Q_{\sigma' b}\}
 =2\delta_{\sigma\sigma'}\delta_{ab}
       (\mathcal H_W+\sigma\Omega K),\qquad \sigma,\sigma'=\pm1,\quad a,b=1,2.
 \label{eq:witten-N4}
\end{equation}
The central generator $K$ is the angular momentum.
Appendix~\ref{app:witten-checks} displays the unreduced matrix
Hamiltonian and supercharges and verifies the mixed anticommutators.

\subsection{The second-order charge between the zero- and two-fermion sectors}

In the component order $(\Psi_0,\Psi_x,\Psi_\varphi,\Psi_2)^t$, the
projectors onto the two scalar sectors are
\begin{equation}
 \Pi_0=\diag(1,0,0,0),\qquad \Pi_2=\diag(0,0,0,1).
\end{equation}
The nonvanishing two-step route is
\begin{equation}
 \Psi_0\ \xrightarrow{\,d_\chi^c\,}\
 \begin{pmatrix}-Y\Psi_0\\X\Psi_0\end{pmatrix}
 \ \xrightarrow{\,d_\chi\,}\ (Y^2+TX)\Psi_0.
\end{equation}
Both steps commute with $\mathcal H_W$, so the product does also.
The projectors retain precisely this transition:
\begin{equation}
 \mathcal Q^{(2)}_+=\tfrac12\Pi_2d_\chi d_\chi^c\Pi_0,\qquad
 \mathcal Q^{(2)}_+\big|_{0\to2}=\tfrac12(Y^2+TX),\qquad
 \mathcal Q^{(2)}_-=(\mathcal Q^{(2)}_+)^\dagger.
 \label{eq:descendant}
\end{equation}
The plus charge sends $\Psi_0$ to the two-fermion sector and annihilates
$\Psi_2$; the minus charge reverses this action. Thus both square to zero.
Order the retained amplitudes as $(\Psi_2,\Psi_0)^t$ and assign them
the grading $\Gamma_{\rm red}=\diag(1,-1)$. They are the plus and minus partner sectors,
respectively. The product raises fermion degree by two and is even
under the original parity. It becomes an odd supercharge under the
new grading, which assigns opposite signs to the retained sectors.

Their surface Hamiltonians, with $W=\Omega F$, are
\begin{equation}
 2\mathcal H_W^{(0,2)}
 =-\hbar^2F^{-1}\partial_x(F\partial_x)
   -\frac{\hbar^2}{F^2}\partial_\varphi^2+W^2\mp2\hbar W'.
 \label{eq:endpoints}
\end{equation}
The upper minus sign belongs to degree zero; the lower plus sign belongs
to degree two. After angular reduction and a common energy shift $a$,
these become $H_-$ and $H_+$.
On the two retained sectors, the composite algebra is
\begin{equation}
 \{\mathcal Q^{(2)}_+,\mathcal Q^{(2)}_-\}
 =\mathcal H_W^2-\Omega^2K^2.
\end{equation}
Appendix~\ref{app:surface-composite} derives this identity from the
first-order brackets. Fixing $K$ will turn its central term into $C$.

\subsection{Fixing angular momentum and changing the radial measure}

Because $K$ is conserved, choose a single integer angular mode $\ell$:
\begin{equation}
 \Psi_j(x,\varphi)=\frac{e^{i\ell\varphi}}{\sqrt{2\pi}}f_j(x),
 \qquad K\Psi_j=k\Psi_j,\qquad k=\hbar\ell,\qquad j=0,2.
\end{equation}
Thus $f_0,f_2$ are the radial amplitudes of the two retained sectors.
Their norms use $F\dd x$. Define the amplitudes in the measure $\dd x$ by
\begin{equation}
 u_-(x)=\sqrt F\,f_0(x),\qquad u_+(x)=\sqrt F\,f_2(x),\qquad
 \int F|f_j|^2\dd x=\int|u_\pm|^2\dd x.
\end{equation}
The multiplication map $T_F:f\mapsto\sqrt F f$ is therefore unitary
between these two radial scalar products, with inverse
$T_F^{-1}:u\mapsto F^{-1/2}u$. In this representation a radial
operator $O$ becomes $T_FOT_F^{-1}$.

Put $g=\hbar F'/(2F)$. Differentiating $F^{-1/2}u$ gives
$T_F(\hbar\partial_x)T_F^{-1}=\hbar\partial_x-g$. Consequently
\begin{equation}
 T_F[-\hbar^2F^{-1}\partial_x(F\partial_x)]T_F^{-1}
 =-\hbar^2\partial_x^2+\Delta_F,\qquad
 \Delta_F=\hbar g'+g^2
 =\hbar^2\left(\frac{F''}{2F}-\frac{F'^2}{4F^2}\right).
 \label{eq:measure}
\end{equation}
The transformed first-order blocks are
\begin{equation}
 A_-=T_FXT_F^{-1}=X-g,\qquad
 A_+=T_FTT_F^{-1}=X+g.
 \label{eq:radial-A}
\end{equation}
Since $[X,g]=\hbar g'$, their product is
$A_+A_-=X^2-[X,g]-g^2=X^2-\Delta_F$.
This commutator accounts for the derivative term and the minus sign
of the correction in the charge. Because $W=\Omega F$,
$\Delta_F=\Delta_W$. Also $Y^2=-k^2/F^2$ on the chosen angular mode.
The reduced operators are therefore
\begin{align}
 H_\pm&=a+\tfrac12[-\hbar^2\partial_x^2+W^2+\Delta_W
                      +k^2/F^2\pm2\hbar W'],\nonumber\\
 q_+&=\tfrac12[(\hbar\partial_x+W)^2-\Delta_W-k^2/F^2],
 \qquad C=-\Omega^2k^2.
 \label{eq:witten-reduced}
\end{align}
Thus $k^2/F^2=-C/W^2$: the cyclic kinetic energy supplies the inverse
superpotential term, while the radial measure supplies the correlated
quantum corrections in $H_\pm$ and $q_+$.

Explicitly, the map from a one-dimensional partner wavefunction to the
surface amplitudes is
\begin{equation}
 J_\ell\begin{pmatrix}u_+\\u_-\end{pmatrix}
 =\frac{e^{i\ell\varphi}}{\sqrt{2\pi F}}
       \begin{pmatrix}u_+\\u_-\end{pmatrix}
 =\begin{pmatrix}\Psi_2\\\Psi_0\end{pmatrix}.
 \label{eq:embedding}
\end{equation}
Restrict $\mathcal Q^{(2)}_\pm$ to the ordered endpoint pair
$(\Psi_2,\Psi_0)^t$, and put
$\mathcal H_{\rm end}=\diag(\mathcal H_W^{(2)}+a,\mathcal H_W^{(0)}+a)$.
Then
\begin{equation}
 \mathcal H_{\rm end}J_\ell=J_\ell\mathbb H,\qquad
 \mathcal Q^{(2)}_\pm J_\ell=J_\ell\mathbb Q_\pm.
 \label{eq:reduction}
\end{equation}
These identities hold on arbitrary smooth $u_+,u_-$.
The resulting system has $n=2$ and $N=2$: one nilpotent charge and
its adjoint, or two Hermitian charges. Its superpartners retain the
zero- and two-fermion sectors of the $N=4$ surface system.

For specified $F,\Omega$, one has
$C_\ell=-\hbar^2\Omega^2\ell^2$, $\ell\in\mathbb Z$.
The distinct values are $0,-\hbar^2\Omega^2,-4\hbar^2\Omega^2,\ldots$;
opposite angular momenta label different sectors with the same invariant.
To reproduce a prescribed nonvanishing $W$ and an arbitrary $C<0$,
one can instead choose $\ell=1$,
$\Omega=\operatorname{sgn}(W)\sqrt{-C}/\hbar$, and $F=W/\Omega$.
For $C=0$, choose $\ell=0$. Thus arbitrary negative $C$ is obtained by
choosing the surface and interaction appropriately, whereas one specified
surface and interaction give the discrete sequence above. At fixed
$W,C$, the choice with $\ell=1$ depends on $\hbar$ through both
$\Omega$ and $F$. A classical limit at fixed geometry and nonzero
$k=\hbar\ell$ instead requires $|\ell|\to\infty$ as $\hbar\to0$.

Finally, for reduced energy $E$ and corresponding surface energy
$E-a$, positivity of Eq.~\eqref{eq:witten-N4} implies
\begin{equation}
 \mathcal H_W+\Omega K\geq0,\qquad
 \mathcal H_W-\Omega K\geq0,\qquad
 E-a\geq|\Omega k|=\sqrt{-C}.
 \label{eq:parent-energy-bound}
\end{equation}
This spectral statement assumes domains on which the full parent
supersymmetry holds. A physical reduction from it can realize only the
upper candidate singlet energy for $C<0$. In particular, reproducing the
full-line oscillator's local expressions does not also supply its lower
singlet: the radial domain matters. The examples below display this
distinction explicitly.

\subsection{Isothermal coordinates and the quantum correction}
\label{sec:ordering}

The auxiliary Hill equation has a direct surface interpretation.
Introduce the dimensionless isothermal coordinate $s$ and the coordinate
$z_W$ adapted to $W$, used in Ref.~\cite{Plyushchay2017}, Eq.~(2.35):
\begin{equation}
 \frac{\dd s}{\dd x}=\frac1F,\qquad
 \frac{\dd z_W}{\dd x}=\frac1W,\qquad
 s=\Omega z_W,\qquad
 \dd\ell_g^2=F^2(\dd s^2+\dd\varphi^2).
 \label{eq:isothermal}
\end{equation}
Additive constants are understood, and $\dd\ell_g^2$ denotes the
surface line element. The metric is locally conformal to the flat cylinder,
and its scalar Laplacian is
$\Delta_g=F^{-2}(\partial_s^2+\partial_\varphi^2)$.
The scalar coefficient of the $0\to2$ composite in
Eq.~\eqref{eq:descendant} is
\begin{equation}
 \mathcal P:=\tfrac12(TX+Y^2)
   =\frac{\hbar^2}{2}e^{-\chi/\hbar}\Delta_g e^{\chi/\hbar}.
 \label{eq:surface-harmonic}
\end{equation}
Thus $\mathcal P\Psi_0=0$ precisely when $e^{\chi/\hbar}\Psi_0$ is harmonic,
locally. For a mode $e^{ik\varphi/\hbar}v(s)$ this is
$v_{ss}-(k/\hbar)^2v=0$. Setting $\psi=\sqrt F\,v(s)$ gives
\begin{equation}
 \mathscr L\psi=F^{-3/2}
       \left(v_{ss}-\frac{k^2}{\hbar^2}v\right),\qquad
 \mathfrak D_W\psi=\Omega\sqrt F\,v_s.
 \label{eq:isothermal-hill}
\end{equation}
Indeed, $U=\Delta_W+k^2/F^2$ in this reduction. Since
$\chi'=\mathcal S'=W$, the reduced kernel function is
$u_-=e^{-\chi/\hbar}\psi$, in agreement with
Eq.~\eqref{eq:hill-charge-main}. The factor $\sqrt F$ also implements
the unitary radial-measure map. The exponential dressing is nonunitary:
normalizability of $v$ uses the measure $F^2e^{-2\chi/\hbar}\dd s$,
with $F,\chi$ evaluated at $x(s)$. The correspondence concerns charge
kernels rather than the full Hamiltonian.

In the $s$ coordinate, the selected vector field $W\partial_x$ is
$\Omega\partial_s$. The full local stabilizer coefficients satisfy
$b_{sss}-4(k/\hbar)^2b_s=0$, with basis
$1,e^{2ks/\hbar},e^{-2ks/\hbar}$ for $k\ne0$, or $1,s,s^2$
for $k=0$. These coefficients $b(s)$ act through
$b\partial_s-b_s/2$. Their density actions form the auxiliary
$\mathfrak{sl}(2,\mathbb R)$; only the selected constant coefficient
has been identified with the gradient in Eq.~\eqref{eq:conformal-gradient}.
These auxiliary actions are not, in general, conserved symmetries of
the full Witten Hamiltonian.

The isothermal coordinate also explains the Schwarzian form:
\begin{equation}
 \{z_W;x\}=\{s;x\}=-\frac{W''}{W}+\frac{W'^2}{2W^2},\qquad
 \Delta_W=-\frac{\hbar^2}{2}\{s;x\}.
 \label{eq:isothermal-schwarzian}
\end{equation}
The first-order operators in Eq.~\eqref{eq:radial-A} are those of
Ref.~\cite{Plyushchay2017}, Eq.~(2.20). With $R_W=\sqrt{|W|}$,
\begin{equation}
 A_-=R_WXR_W^{-1},\qquad A_+=R_W^{-1}XR_W,\qquad
 W^{-1}A_-=A_+W^{-1}.
 \label{eq:radial-connections}
\end{equation}
Their opposite logarithmic terms originate in the surface measure and
frame; the identity $W(A_--W)=\hbar\mathfrak D_W$ relates one of
them directly to the weight $-1/2$ Lie derivative. The multiplier
$1/W$ intertwines these two first-order operators. For $C\ne0$,
the scalar Darboux factors additionally contain $\pm\beta/W$,
$\beta^2=-C$, as in Eq.~\eqref{eq:beta-factors}.

This also fixes the data of the quantization prescription in
Ref.~\cite{Plyushchay2017}, Eq.~(2.41). The classical identity
$p^2=\zeta p\zeta^{-2}p\zeta$, developed in
Ref.~\cite{BravoPlyushchay2016}, Sections~2 and~3, allows arbitrary
$\zeta$. Keeping this order after quantization produces
$D_\zeta^\dagger D_\zeta$, with
$D_\zeta=\hbar\zeta^{-1}\partial_x\zeta$.
The surface measure selects $\zeta\propto F^{-1/2}\propto|W|^{-1/2}$,
so Eq.~\eqref{eq:measure} gives the required corrected kinetic operator;
$A_+A_-$ gives the opposite correction in the charge.
The original flat kinetic energy has no ordering ambiguity: the extra
data come from quantization of a specified surface followed by reduction.

Nor is $\Delta_W$ determined by Gaussian curvature alone.
The radius functions $F_1=Re^{\kappa x}$ and
$F_2=R\cosh(\kappa x)$ have the same curvature $-\kappa^2$, but
\begin{equation}
 \Delta_{F_1}=\frac{\hbar^2\kappa^2}{4},\qquad
 \Delta_{F_2}=\frac{\hbar^2\kappa^2}{4}
                 [1+\operatorname{sech}^2(\kappa x)].
 \label{eq:equal-curvature-corrections}
\end{equation}
They describe different cyclic Killing orbits on the same local geometry.
The conversion of cyclic kinetic energy into an inverse squared radius
potential also underlies the finite-gap reductions in
Ref.~\cite{BravoPlyushchay2016}, Section~4.7, and the AdS$_2$
Aharonov--Bohm construction of Ref.~\cite{CorreaJakubskyPlyushchay2009}.
Here $W=\Omega F$ additionally ensures conservation of the parent
charges and the complete reduced superalgebra.

\section{Metric signature and examples}
\label{sec:signature}

The second-order symmetry obtained above acts between the zero-fermion
and two-fermion sectors. The sign of their cyclic kinetic term determines
$C$. We now reverse that sign, keeping track of the fermion metric and
the adjoint, and illustrate the resulting operators.

\subsection{The indefinite surface and its fermions}

Replace the circle by a noncompact cyclic coordinate $y\in\mathbb R$,
and take
\begin{equation}
 \dd s_\varepsilon^2=\dd x^2+\varepsilon F(x)^2\dd y^2,
 \qquad \varepsilon=\pm1,\qquad F>0,\qquad W=\Omega F.
 \label{eq:signed-metric}
\end{equation}
Both $y$ and the earlier angle are dimensionless. For $\varepsilon=-1$
the surface is Lorentzian; $y$ is a configuration coordinate, not the
Schr\"odinger evolution time. The volume density is still $F\dd x\dd y$.

The metric induces signs on the fermionic states. With the same ordering
$(\Psi_0,\Psi_x,\Psi_y,\Psi_2)^t$ as in Section~\ref{sec:witten},
\begin{equation}
 G_\varepsilon=\diag(1,1,\varepsilon,\varepsilon),\qquad
 \langle\Psi,\Phi\rangle_\varepsilon
 =\int F\dd x\dd y\,\Psi^\dagger G_\varepsilon\Phi.
 \label{eq:signed-pairing}
\end{equation}
Let $f_i^+$ create a fermion in direction $i=x,y$, and let
$f_i^-=(f_i^+)^\sharp$ be its adjoint in this pairing. Then
\begin{equation}
 \{f_x^-,f_x^+\}=1,\qquad \{f_y^-,f_y^+\}=\varepsilon,
 \qquad \{f_x^\pm,f_y^\pm\}=\{f_x^\pm,f_y^\mp\}=0.
\end{equation}
For $\varepsilon=-1$, the $y$ fermionic mode has a negative
anticommutator. States containing that fermion have negative norm, including
the two-fermion state. The adjoint $\sharp$ therefore differs from the
positive componentwise adjoint $\dagger$:
$A^\sharp=G_\varepsilon^{-1}A^\dagger G_\varepsilon$.
The latter uses the same measure but assigns positive norms to all four
components.

To display the first-order maps, set
$X=\hbar\partial_x+W$, $T=X+\hbar F'/F$ and
$Y=\hbar F^{-1}\partial_y$. The differential $d_\chi$ acts as
\begin{equation}
 \Psi_0\longmapsto\begin{pmatrix}X\Psi_0\\Y\Psi_0\end{pmatrix},
 \qquad
 \begin{pmatrix}\Psi_x\\\Psi_y\end{pmatrix}
       \longmapsto -Y\Psi_x+T\Psi_y.
\end{equation}
Define its oriented companion by the matrix
\begin{equation}
 \mathcal R_\varepsilon=
 \begin{pmatrix}0&-\varepsilon\\1&0\end{pmatrix}
 \quad\hbox{on one-fermion components},
 \qquad \mathcal R_\varepsilon^2=-\varepsilon\id_2.
\end{equation}
Extend this map as $1$ on degree zero and $\varepsilon$ on degree two.
Then $d^c_{\chi,\varepsilon}=\mathcal R_\varepsilon d_\chi
\mathcal R_\varepsilon^{-1}$ acts by
\begin{equation}
 \Psi_0\longmapsto\begin{pmatrix}-\varepsilon Y\Psi_0\\X\Psi_0\end{pmatrix},
 \qquad
 \begin{pmatrix}\Psi_x\\\Psi_y\end{pmatrix}
       \longmapsto -T\Psi_x-\varepsilon Y\Psi_y.
\end{equation}
For positive signature, $\mathcal R_+$ is the quarter-turn used before.
For negative signature, $\mathcal R_-^t\diag(1,-1)
+\diag(1,-1)\mathcal R_-=0$: it is a Lorentz-boost generator.
The finite boost is
$e^{\tau\mathcal R_-}=\cosh\tau\,\id_2+\sinh\tau\,\mathcal R_-$.
The matrix $\mathcal R_-$ itself exchanges the two components and reverses
their norm sign; it is not that finite boost.

With $Q=-id_\chi$, $\widetilde Q=d^c_{\chi,\varepsilon}$,
$K=-i\hbar\partial_y$ and
$\mathcal H_\varepsilon=\{Q,Q^\sharp\}/2$, one obtains
\begin{align}
 Q^2&=\widetilde Q^2=\{Q,\widetilde Q\}=0,\nonumber\\
 \{Q,Q^\sharp\}&=2\mathcal H_\varepsilon,\qquad
 \{\widetilde Q,\widetilde Q^\sharp\}=2\varepsilon\mathcal H_\varepsilon,
 \nonumber\\
 \{Q,\widetilde Q^\sharp\}&=\{\widetilde Q,Q^\sharp\}=2\varepsilon\Omega K.
 \label{eq:signed-algebra}
\end{align}
All four supercharges commute with $\mathcal H_\varepsilon$ and $K$.
Their unreduced matrices and signed-adjoint prescription are given in
Appendix~\ref{app:witten-checks}.
For negative signature this algebra uses an indefinite adjoint, so its
anticommutators no longer imply the positivity bound of
Eq.~\eqref{eq:parent-energy-bound}.
The degree-zero to degree-two composite has coefficient
$(TX+\varepsilon Y^2)/2$.

\subsection{Reduction at fixed cyclic momentum}

Translation invariance in $y$ allows a Fourier decomposition. For either
retained component $j=0,2$, write
\begin{equation}
 \Psi_j(x,y)=\frac1{\sqrt{2\pi\hbar}}
       \int_{\mathbb R}\dd k\,e^{iky/\hbar}f_{j,k}(x),
 \qquad K e^{iky/\hbar}=k e^{iky/\hbar}.
\end{equation}
The real number $k$ is the conserved momentum conjugate to $y$ and
has dimensions of action. Since the coefficients do not depend on $y$, each
Fourier component is governed by a one-dimensional operator. In that component
$Y^2=-k^2/F^2$. For a periodic coordinate of period $2\pi$, the integral
is instead a Fourier series and $k=\hbar\ell$, $\ell\in\mathbb Z$.
For noncompact $y$, a fixed $k$ labels a fiber of the direct-integral
Fourier decomposition, not a subspace of normalizable plane waves.
Square-integrability of the reduced $x$-state is therefore a statement
within that fiber. This applies also to the noncompact Pauli reductions
below; in the indefinite model the component pairing remains signed.

For each $k$, set $u_- =\sqrt F f_{0,k}$ and
$u_+=\sqrt F f_{2,k}$. This removes the radial measure while preserving
the inherited component signs: multiplication by $\sqrt F$ does not
turn the negative two-form norm into a positive one.
The resulting scalar operators are
\begin{align}
 H_{\varepsilon,\pm}
 &=a+\tfrac12[-\hbar^2\partial_x^2+W^2+\Delta_W
                    +\varepsilon k^2/F^2\pm2\hbar W'],\nonumber\\
 q_{+,\varepsilon}
 &=\tfrac12[(\hbar\partial_x+W)^2-\Delta_W-\varepsilon k^2/F^2],
 \qquad C=-\varepsilon\Omega^2k^2.
 \label{eq:signed-reduced}
\end{align}
The volume density produces the same correction for both signatures.
Only the cyclic kinetic term changes sign. Thus nonzero $k$ gives
$C<0$ for $\varepsilon=+1$ and $C>0$ for $\varepsilon=-1$;
$k=0$ gives $C=0$ in both.
In the isothermal coordinate of Eq.~\eqref{eq:isothermal}, the same
calculation gives the separated equation
$v_{ss}-\varepsilon(k/\hbar)^2v=0$ and the Bol equation
$b_{sss}-4\varepsilon(k/\hbar)^2b_s=0$.
For $\varepsilon=-1$ and $k\ne0$, the Hill solutions are
trigonometric, while the Bol solution space is spanned by
$1,\cos(2ks/\hbar),\sin(2ks/\hbar)$. The resulting
elliptic flow remains an element of the auxiliary $SO^+(2,1)$
representation described in Section~\ref{sec:target}; changing the
surface signature does not change that auxiliary group.

The final step defines the ordinary positive inner product on the two
retained scalar sectors. The inherited reverse coefficient is
$\varepsilon q_{+,\varepsilon}^\dagger$; using the positive adjoint
$q_-=q_+^\dagger$ therefore multiplies it by $\varepsilon$ and yields
Eq.~\eqref{eq:poly} with the displayed $C$. For $\varepsilon=-1$, this
choice changes the pairing. Fourier reduction and the radial-density
map alone do not make this a unitary reduction of an indefinite parent
to a positive Hilbert space.
The relation to the non-Hermitian scalar intermediate in
Ref.~\cite{Plyushchay2017} is derived in Section~\ref{sec:limits}:
that intermediate is one of the two scalar channels of this one-fermion
matrix Hamiltonian.

\subsection{Cones, angular deficit and the radial oscillator}
\label{sec:cone}

Take $x=r>0$, $F=\gamma r$ and $W=br$, where
$\gamma>0$ and $b=\Omega\gamma\ne0$. In positive signature with a
$2\pi$-periodic angle, this is an intrinsic cone: a circle of proper
radius $r$ has circumference $2\pi\gamma r$. The angular deficit is
$2\pi(1-\gamma)$ for $0<\gamma<1$; $\gamma>1$ gives angular
excess $2\pi(\gamma-1)$, while $\gamma=1$ gives the plane.
Here $k=\hbar\ell$ is angular momentum.
For the negative signature with noncompact $y$, the metric
$\dd r^2-\gamma^2r^2\dd y^2$ is a Rindler wedge of flat Minkowski
space; $k$ is the cyclic momentum defined in the preceding subsection,
not a quantized angular momentum. The Euclidean deficit-angle
interpretation does not apply to this noncompact Lorentzian case.

In either case
\begin{equation}
 \Delta_W=-\frac{\hbar^2}{4r^2},\qquad
 C=-\varepsilon\frac{b^2k^2}{\gamma^2},\qquad
 g_\varepsilon=\frac{\varepsilon k^2}{\gamma^2}-\frac{\hbar^2}{4},
 \label{eq:cone-parameters}
\end{equation}
and
\begin{align}
 H_{\varepsilon,\pm}
 &=a+\frac12\left[-\hbar^2\partial_r^2+b^2r^2
                         +\frac{g_\varepsilon}{r^2}\right]\pm\hbar b,
 \label{eq:cone-example}\\
 q_+&=\frac12\left[(\hbar\partial_r+br)^2-\frac{g_\varepsilon}{r^2}\right].
\end{align}
These are supersymmetric partners of the oscillator form of AFF
conformal mechanics \cite{deAlfaroFubiniFurlan1976}. The inverse-square
coupling is the sum of the cyclic contribution and the radial quantum
term $-\hbar^2/4$.

The positive cone connects this construction directly with the oscillator
in a cosmic string background
\cite{InzunzaPlyushchayCone2021,InzunzaPlyushchayRotatingCone2022}.
Their spatial cone metric is $\dd s^2=\alpha^2\dd\rho^2+\rho^2\dd\varphi^2$;
the correspondence is $r=\alpha\rho$ and $\gamma=1/\alpha$.
In the rotating model of Ref.~\cite{InzunzaPlyushchayRotatingCone2022},
the added term is $-\alpha\Omega_{\rm rot}K$, with
$K=-i\hbar\partial_\varphi$ and $\Omega_{\rm rot}$ the rotation
parameter of that reference. At fixed angular momentum
$k=\hbar\ell$, adding it to both partners shifts
$a\mapsto a-\alpha\Omega_{\rm rot}\hbar\ell$ and leaves
$q_\pm$, $g_+$ and $C$ unchanged. The spatial metric remains
Riemannian; rotation therefore does not implement the change of
signature considered here.

To identify the radial operators, take $b>0$, set $g=g_+$ and denote the
common AFF Hamiltonian by
\begin{equation}
 h_g=\frac12\left[-\hbar^2\partial_r^2+b^2r^2+\frac{g}{r^2}\right],
 \qquad H_\pm=a+h_g\pm\hbar b.
 \label{eq:cone-aff-hamiltonian}
\end{equation}
The same $q_\pm$ that intertwine $H_-$ and $H_+$ are lowering and
raising operators for $h_g$:
\begin{equation}
 [h_g,q_+]=-2\hbar b\,q_+,\qquad
 [h_g,q_-]=2\hbar b\,q_-,\qquad q_-=q_+^\dagger.
 \label{eq:cone-aff-ladders}
\end{equation}
Indeed, the first identity gives
$(a+h_g+\hbar b)q_+=q_+(a+h_g-\hbar b)$, which is precisely
$H_+q_+=q_+H_-$. The relative energy shift of the partners thus
compensates the energy change produced by the radial ladder operator.
In the matrix system this makes the associated supercharge conserved.
The dimensionless operators
\begin{equation}
 J_0=\frac{h_g}{2\hbar b},\qquad
 J_-=\frac{q_+}{2\hbar b},\qquad J_+=\frac{q_-}{2\hbar b}
 \label{eq:cone-aff-generators}
\end{equation}
satisfy $[J_0,J_\pm]=\pm J_\pm$ and $[J_-,J_+]=2J_0$:
our second-order intertwiner is a generator of the radial conformal
$\mathfrak{sl}(2,\mathbb R)$ algebra. This is a simple instance of the
relation between constant shifts of Darboux partners and ladder
operators discussed in Ref.~\cite{PlyushchayExotic2020}. The ladder
relations do not themselves imply an additional stationary Lax--Novikov
integral or an extra pair of supercharges.

For the positive cone, put $\nu=|\ell|/\gamma$, so that
$g=\hbar^2(\nu^2-1/4)$. Near the tip the two local scalar solutions
behave as $r^{1/2\pm\nu}$, or as $\sqrt r$ and
$\sqrt r\log(r/r_0)$ when $\nu=0$. Here $r_0>0$ is a reference
length. Both solutions lie in $L^2(\dd r)$ for $0\leq\nu<1$,
so a boundary condition is needed (a limit-circle endpoint); for
$\nu\geq1$, only the plus branch is square-integrable (a limit-point
endpoint) \cite{GitmanTyutinVoronov2010}. At $\nu=1/2$, the
inverse-square coupling vanishes, but a half-line boundary condition
is still needed. The regular oscillator expressions then apply. For
example, $b>0$, $\gamma=2$, $|\ell|=1$ and the radial boundary
condition $u(0)=0$ retain the odd full-line oscillator states. The lowest
$H_-$ level is $a+\hbar b/2$, from $|1\rangle$; the lower full-line
singlet $|0\rangle$ violates that boundary condition. At $\ell=0$,
instead, $C=0$ and $g=-\hbar^2/4$. In negative signature and
$k\ne0$, $g<-\hbar^2/4$ is the strongly attractive inverse-square
regime; this particular singular example has no lower energy bound.

The cancellation at $\gamma=2$ requires an angular excess and cannot
occur at an ordinary smooth polar axis in this scalar Witten reduction.
With angular period $2\pi$, a smooth axis has
$F(r)=r+O(r^3)$ and integer $\ell$ in both endpoint sectors. Since
$W=\Omega F$,
\begin{equation}
 U(r)=\frac{\hbar^2(\ell^2-1/4)}{r^2}+O(1),\qquad
 C=-\hbar^2\Omega^2\ell^2.
 \label{eq:smooth-axis-obstruction}
\end{equation}
Smooth continuation of $U$ across the simple zero would require
$C=-\hbar^2\Omega^2/4$, or $\ell^2=1/4$, which is impossible here.
Thus a smooth two-dimensional parent need not yield smooth full-line
one-dimensional coefficients. Its half-line radial realization can
nevertheless be well defined. The preceding classification now has
$\nu=|\ell|$: only $\ell=0$ is limit-circle, and regularity of the
parent amplitude $u/\sqrt F$ selects its nonlogarithmic branch.
For $|\ell|\geq1$, the endpoint is limit-point. This scalar
classification does not specify common domains for all four surface charges.

For the positive cone, the radial relations
\eqref{eq:cone-aff-ladders} hold as differential identities for every
$\gamma>0$; their physical realization requires compatible radial
operator domains. The additional angular hidden symmetries of
Refs.~\cite{InzunzaPlyushchayCone2021,InzunzaPlyushchayRotatingCone2022}
raise a different issue. For the nonrotating oscillator, globally
defined classical hidden integrals exist at rational $\alpha$, whereas
the corresponding quantum angular operators constructed there preserve
the physical state space only at integer $\alpha$. Rational noninteger
$\alpha$ therefore exhibits a geometric quantum anomaly. Unlike
$q_\pm$, these operators change angular momentum and generally connect
different reduced couplings $g$ and invariants $C$.
Our example $\gamma=2$, $|\ell|=1$ has $\alpha=1/2$: the finite span
of odd oscillator eigenstates is an invariant core for the radial
ladders, despite the obstruction to those angular symmetries.
The distinction between this invariant core and the full Dirichlet
Hamiltonian domain is illustrated in Appendix~\ref{app:quantum-domains};
domains for the unreduced first-order charges still require separate
analysis. Thus the correction
$\Delta_W$ that restores the local second-order SUSY identities does
not by itself resolve the global domain obstruction for the larger
angular symmetry algebra.

\subsection{A smooth non-affine example and a known special case}

Choose
\begin{equation}
 F(x)=R\cosh(\kappa x),\qquad W(x)=b\cosh(\kappa x),
 \qquad R>0,\quad\kappa>0,\quad b=\Omega R\ne0.
\end{equation}
The metric scale $F(x)$ is bounded below by the constant length $R$;
in the periodic Riemannian case it is the local circle radius. Here
$\kappa$ has dimensions of inverse length. The correction and invariant are
\begin{equation}
 \Delta_W=\frac{\hbar^2\kappa^2}{4}
               [1+\operatorname{sech}^2(\kappa x)],\qquad
 C=-\varepsilon\frac{b^2k^2}{R^2},
\end{equation}
where $k$ retains its meaning as cyclic momentum. The partner operators
become
\begin{equation}
 \begin{aligned}
 2(H_{\varepsilon,\pm}-a)
 ={}&-\hbar^2\partial_x^2+b^2\cosh^2(\kappa x)
                         +\frac{\hbar^2\kappa^2}{4}\\
 &+\left(\frac{\hbar^2\kappa^2}{4}-\frac C{b^2}\right)
             \operatorname{sech}^2(\kappa x)
       \pm2\hbar b\kappa\sinh(\kappa x).
 \end{aligned}
 \label{eq:cosh-example}
\end{equation}
Both potentials are smooth and confining on the real line for either
sign of $C$: the positive $\cosh^2$ term dominates the opposite
supersymmetric contributions $\pm2\hbar W'$ at both ends. In particular,
$C>0$ does not itself imply a singular reduced potential.
In this Witten construction, the $C>0$ members come from the
indefinite parent; smooth
scalar potentials in the reduced positive Hilbert space do not change
the parent's pairing.

The family contains a known member of the exponential-superpotential
class of Ref.~\cite{KlishevichPlyushchay2001}, Section~3.2,
Eqs.~(3.9)--(3.11) of that reference. Namely, for
$C=\hbar^2\kappa^2b^2/4$, the $\operatorname{sech}^2$ term cancels,
$U=\hbar^2\kappa^2/4$, and
\begin{equation}
 q_+=\frac12\left[(\hbar\partial_x+b\cosh\kappa x)^2
                                  -\frac{\hbar^2\kappa^2}{4}\right].
\end{equation}
This is the $n=2$ specialization with equal exponential coefficients
$w_+=w_-=b/2$ and $w_0=0$; their additive energy constant is
$v=a+\hbar^2\kappa^2/8$. Generic $C$ retains the additional
$\operatorname{sech}^2$ term. We use the resulting family as a smooth
non-affine example of the reduction, without assuming exact solvability.

For $\varepsilon=-1$, $R=L$ and $\kappa=L^{-1}$, the same
surface is the universal cover of AdS$_2$ when $x,y\in\mathbb R$.
Its isothermal coordinate and metric are
\begin{equation}
 s=\arctan[\sinh(x/L)],\qquad
 \dd\ell_g^2=L^2\sec^2s\,(\dd s^2-\dd y^2),\qquad
 |s|<\pi/2.
 \label{eq:ads-isothermal}
\end{equation}
Here $\chi=\Omega L^2\sinh(x/L)$, up to a constant.
The signed version of Eq.~\eqref{eq:surface-harmonic} is
$\mathcal P_-=(TX-Y^2)/2=(\hbar^2/2)e^{-\chi/\hbar}\Box_g e^{\chi/\hbar}$,
with $\Box_g$ the scalar wave operator. Its local kernel therefore maps
to the massless AdS$_2$ equation; a mode $e^{iky/\hbar}v(s)$ obeys
$v_{ss}+(k/\hbar)^2v=0$. This is a concrete connection with
geometrizations based on the AdS Klein--Gordon equation, including the
Higgs-oscillator constructions of
Refs.~\cite{EvninNivesvivat2016,EvninNivesvivat2017}.
It concerns the composite-charge kernel, not the full physical
Hamiltonian. Moreover, AdS isometries generally change $k$, whereas
the auxiliary Hill stabilizer acts within a fixed-mode solution space.
The two $\mathfrak{sl}(2,\mathbb R)$ actions must therefore be distinguished.

\section{Pauli reductions and their relation to the surface construction}
\label{sec:pauli}

The preceding construction combines two fermionic first-order operators.
There is another possibility: combine one fermionic supercharge with a
bosonic first-order symmetry. Their product is still fermionic under the
original grading, so no change of grading is required. A charged spin-$1/2$
particle in a uniform magnetic field on the flat plane provides a concrete example. Its
additional first-order symmetries describe the guiding centre of the
cyclotron motion \cite{KimLee2002,InzunzaPlyushchay2022}.

\subsection{A first-order supercharge and a spatial symmetry}

Use polar coordinates on the plane,
$\dd s^2=\dd r^2+r^2\dd\varphi^2$, and choose the magnetic potential
$A=\omega r^2\dd\varphi$. The electric charge is absorbed into $A$;
the magnetic field is $B=2\omega$. We express the two spin components
in a frame that rotates with the polar coordinates. The inner product
uses the measure $r\dd r\dd\varphi$, and
\begin{equation}
 J=-i\hbar\partial_\varphi,\qquad
 \delta_r=\partial_r+\frac1{2r},\qquad
 \Gamma=\sigma_3,\qquad \sigma_+=\frac{\sigma_1+i\sigma_2}{2}.
 \label{eq:pauli-frame}
\end{equation}
The term $1/(2r)$ makes $\delta_r$ formally anti-Hermitian in the
radial measure. The Pauli supercharges and Hamiltonian are
\begin{equation}
 Q_+=\sigma_+\left(\hbar\delta_r+\frac Jr-\omega r\right),\qquad
 Q_-=Q_+^\dagger,\qquad
 \mathcal H_P=\tfrac12\{Q_+,Q_-\}.
 \label{eq:pauli-first}
\end{equation}
Here $Q_+$ changes the spin label from $-$ to $+$ while preserving
the eigenvalue of $J$. These first-order nilpotent operators commute
with $\mathcal H_P$.

The required bosonic symmetry is
\begin{equation}
 R_-=e^{-i\varphi}\left(\hbar\partial_r+\frac Jr+\omega r
                                     -\frac{\hbar\Gamma}{2r}\right),
 \qquad R_+=R_-^\dagger.
 \label{eq:pauli-even}
\end{equation}
It preserves the spin label, lowers the eigenvalue of $J$ by $\hbar$,
and commutes with both $\mathcal H_P$ and $Q_\pm$. Thus $Q_+R_-$
changes spin and angular momentum together. The combination left
unchanged is
\begin{equation}
 K=J+\frac\hbar2\Gamma.
\end{equation}
Indeed, $[K,Q_+]=\hbar Q_+$ and $[K,R_-]=-\hbar R_-$.
Consequently the second-order operators
\begin{equation}
 S_+=Q_+R_-,\qquad S_-=S_+^\dagger
\end{equation}
can be restricted to a fixed $K=k$ sector. Their algebra is
\begin{equation}
 [K,S_\pm]=0,\qquad
 \{S_+,S_-\}=2\mathcal H_P(2\mathcal H_P+4\omega K).
 \label{eq:pauli-composite}
\end{equation}
The first-order steps and this polynomial identity are checked in
Appendix~\ref{app:pauli}.

To perform the reduction, label a state by $(s,j)$, where $s=\pm1$
is its $\Gamma$ eigenvalue and $j$ its $J$ eigenvalue. The action of $S_+$
within $K=k$ passes through the following three spaces:
\begin{equation}
 (-,k+\hbar/2)\ \xrightarrow{R_-}\
 (-,k-\hbar/2)\ \xrightarrow{Q_+}\ (+,k-\hbar/2).
 \label{eq:pauli-route}
\end{equation}
The intermediate state has a different value of $K$; only the
\emph{product} of the two steps preserves $K$. This is why one must
keep the angular shift when extracting the radial operators.

Write the wavefunction in each channel as
$e^{ij\varphi/\hbar}u(r)e_s/\sqrt{2\pi r}$, where
$\Gamma e_s=se_s$. The radial inner product then becomes
$\int_0^\infty |u|^2\dd r$. Substitution into the two steps of
Eq.~\eqref{eq:pauli-route} gives
\begin{equation}
 T=\hbar\partial_r+\frac{k+\hbar/2}{r}+\omega r,\qquad
 V=\hbar\partial_r+\frac{k-\hbar/2}{r}-\omega r,\qquad
 q_+=\tfrac12VT.
 \label{eq:pauli-legs}
\end{equation}
The factor $1/2$ fixes the normalization to Eq.~\eqref{eq:q}.
Let $h_-$ and $h_+$ denote the radial restrictions of
$\mathcal H_P$ on the first and last channels. Equation~\eqref{eq:pauli-composite}
then implies
\begin{equation}
 q_-q_+=h_-(h_-+2\omega k)=(h_-+\omega k)^2-\omega^2k^2.
\end{equation}
This identity explains the energy shift needed to compare with the
one-dimensional notation: set $H_\pm=h_\pm+a+\omega k$.
The shift is made after fixing $K=k$. Promoting it to $\omega K$ in
the unreduced Hamiltonian would give
\begin{equation}
 [\mathcal H_P+\omega K,Q_+]=\hbar\omega Q_+,\qquad
 [\mathcal H_P+\omega K,R_-]=-\hbar\omega R_-.
 \label{eq:pauli-shift-conservation}
\end{equation}
The product $Q_+R_-$ would remain conserved; for $\omega\ne0$, its
separate first-order factors would not. Thus conservation of these
factors must be checked against the original surface Hamiltonian.

For $k\ne0$, multiplication of $V$ and $T$ now gives the target data
\begin{equation}
 W=\frac kr,\qquad
 \Delta_W=\frac{3\hbar^2}{4r^2},\qquad
 U=\frac{3\hbar^2}{4r^2}+\omega^2r^2,\qquad
 C=-\omega^2k^2,
 \label{eq:pauli-target}
\end{equation}
with partner Hamiltonians
\begin{equation}
 H_\pm=a+\frac12\left[-\hbar^2\partial_r^2+\omega^2r^2
                  +\frac{k^2+3\hbar^2/4\mp2\hbar k}{r^2}\right].
 \label{eq:pauli-radial-H}
\end{equation}
For the spin structure inherited from single-valued Cartesian spinors,
the rotating-frame components are antiperiodic, giving $k=\hbar\ell$,
$\ell\in\mathbb Z$. The boundary conditions at $r=0$ must likewise be
inherited from the planar problem; the differential expressions alone
do not specify them.

The radial operators also cover $k=0$ for every $\omega$. In this sector
$W\equiv0$, $H_+=H_-$ and $q_+=q_-=-(H_\pm-a)$; only the
parametrization involving inverse powers of $W$ is inapplicable.
Since $k=\hbar\ell$ is discrete, this is a separate angular sector
at fixed $\hbar$, rather than a continuous limit through nonzero
allowed sectors. The repeated algebraic root $E=a$ is not a physical
singlet: its formal radial zero mode is not square-integrable at the
axis. Appendix~\ref{app:pauli} proves this and specifies the endpoint
conditions in all angular sectors.

This example gives two useful physical conclusions. First, a nonzero
$\Delta_W$ can arise from a \emph{flat} parent. Here its coefficient
comes from the radial measure, spin transport, and the angular shift.
The correction is therefore not universally proportional to the
surface scalar curvature.
Second, the parent is not unique. For each fixed $\omega>0$ and $k>0$, the same data
\eqref{eq:pauli-target} follow from the Witten construction with
$\Omega=\omega$, $F=k/(\omega r)$, and cyclic momentum $k$.
That surface has Gaussian curvature $-F''/F=-2/r^2$, whereas the
Pauli surface is flat. In the Pauli construction the oscillator term
comes from the magnetic field; in the Witten construction it comes
from the cyclic kinetic energy. These distinct mechanisms yield the
same local radial expressions, without identifying their parent
geometries or global domains.

\subsection{Relation to angular reduction and to nonlinear Pauli systems}
\label{sec:pauli-extensions}

The zero-field limit connects this construction directly to the
angular reduction of Ref.~\cite{AnabalonPlyushchay2003}. Setting
$\omega=0$ gives $W=k/r$, $U=3\hbar^2/(4r^2)$ and $C=0$, with
\begin{equation}
 q_+=\frac12\left(\hbar\partial_r+\frac{k-\hbar/2}{r}\right)
                \left(\hbar\partial_r+\frac{k+\hbar/2}{r}\right).
 \label{eq:pauli-free-limit}
\end{equation}
In units $\hbar=1$, put $k=\alpha-1/2$. These local differential
expressions coincide with its $n=2$ Hamiltonian and stationary
supercharge, up to charge normalization and the energy origin.
The correspondence of the angular constraints is equally direct:
in the Cartesian spin frame, $K=L+\hbar\Gamma$, where $L$ is orbital
angular momentum. Thus fixing $K$ weights the spin twice as strongly
as in the usual total angular momentum $L+\hbar\Gamma/2$.
The general integer weight $n$ in that earlier construction produces
order-$n$ supercharges. Its quantize-then-reduce procedure also explains
the quantum terms that arise from the radial measure and the angular shifts.

For the uniform-field, $g=2$ mechanism used above, real $\omega$ and
$k$ always give $C=-\omega^2k^2\leq0$. Replacing the uniform field
by an arbitrary radial field does not preserve the commuting
guiding-centre operator $R_-$ used in $S_+=Q_+R_-$, even though
the ordinary Pauli supersymmetry survives. Consequently this argument
does not extend to arbitrary $W$ merely by changing the field profile.

A broader Pauli construction can nevertheless give $C>0$ on the
Euclidean plane with a positive-definite Hilbert-space inner product.
Nonlinear Pauli supersymmetry with gyromagnetic
ratio $g=2n$ was studied in Ref.~\cite{KlishevichPlyushchayPlane2001}.
For its $n=2$ exponential-field example, use Cartesian length coordinates
$x,y$ and denote the linear momentum conjugate to $y$ by $k$.
Thus $k$ differs from the angular momentum of the polar construction.
Use $\Pi_i=-i\hbar\partial_i-A_i$ and
$\mathcal H_g=(\Pi_x^2+\Pi_y^2)/2-g\hbar B\Gamma/4$, with
\begin{equation}
 A_x=0,\qquad A_y=-b\cosh(\kappa x),\qquad
 B=-b\kappa\sinh(\kappa x).
\end{equation}
For $g=4$, reduction at $-i\hbar\partial_y=k$ gives
\begin{equation}
 \mathcal H_{4,k}=\frac12[-\hbar^2\partial_x^2+W_k^2]
                  +\hbar W_k'\Gamma,
 \qquad W_k=k+b\cosh(\kappa x).
 \label{eq:nonlinear-pauli-reduced}
\end{equation}
Its scalar intertwiner is
\begin{equation}
 q_{+,k}=\frac12\left[(\hbar\partial_x+W_k)^2
                              -\frac{\hbar^2\kappa^2}{4}\right].
\end{equation}
The corresponding planar supercharge is obtained by restoring the
cyclic momentum operator:
\begin{equation}
 \mathcal S_{4,+}=\frac{\sigma_+}{2}
 \left[\left(\hbar\partial_x-i\hbar\partial_y+b\cosh(\kappa x)\right)^2
                         -\frac{\hbar^2\kappa^2}{4}\right],
 \qquad \mathcal S_{4,-}=\mathcal S_{4,+}^\dagger.
\end{equation}
It is nilpotent and commutes with $\mathcal H_4$.
Indeed, $W_k'''=\kappa^2W_k'$ solves the compatibility equation
with $U=\hbar^2\kappa^2/4$. With the energy origin of
Eq.~\eqref{eq:nonlinear-pauli-reduced}, $a=-\hbar^2\kappa^2/8$,
and the first integral gives
\begin{equation}
 4C_k=\hbar^2[2W_kW_k''-(W_k')^2]-4UW_k^2
     =\hbar^2\kappa^2(b^2-k^2).
 \label{eq:nonlinear-pauli-C}
\end{equation}
Thus $|k|<|b|$ gives $C_k>0$; the $k=0$ Fourier sector gives the known
member of the $\cosh$ family in Eq.~\eqref{eq:cosh-example}. This positive-metric
realization starts with \emph{second-order} odd charges on the plane.
It therefore supplies a useful comparison, but does not furnish
the conserved first-order parent required for a general $C>0$ target.

The same family illustrates regular continuation through zeros
(Appendix~\ref{app:zeros}). For $b,\kappa>0$, $k=-b$ gives an
admissible double zero at $x=0$ with $C_k=0$, whereas $k<-b$ gives
two simple zeros $x_\pm$ with $|W_k'(x_\pm)|=\kappa\sqrt{k^2-b^2}$ and
$C_k=-\hbar^2[W_k'(x_\pm)]^2/4$. In every case, the complete
$U=\hbar^2\kappa^2/4$ is smooth. These sign conditions permit zeros
without requiring them: $k=b$ and $k>b$ give nonvanishing $W_k$
with $C_k=0$ and $C_k<0$, respectively.

\subsection{Curvature and symmetry enhancement under reduction}
\label{sec:hyperbolic-pauli}

The scalar Landau problem on the hyperbolic plane and its
$SL(2,\mathbb R)$ description were studied in Ref.~\cite{Comtet1987}. 
Its relation to the planar Kepler problem through a common Morse
system is developed in Ref.~\cite{Plyushchay2026}.
Factorization of magnetic Schr\"odinger operators on curved surfaces,
including curvature-dependent magnetic-field shifts and global bundle
conditions, was developed in Ref.~\cite{FerapontovVeselov2001}.
Supersymmetric magnetic dynamics on such surfaces is treated in
Refs.~\cite{KimLee2002,KlishevichPlyushchayCurved2002}.
Here two reductions separate two questions: how curvature changes the
quadratic invariant, and when the reduced system has more conserved
charges than those directly inherited from its parent.
We use the intrinsic spinor Pauli Hamiltonian defined by the square
of its first-order Dirac operator. The spin connection is included
in the charges below; a scalar Landau Hamiltonian with a Zeeman term
alone would not specify this model.

First use polar coordinates on the hyperbolic plane of Gaussian
curvature $-1/R^2$:
\begin{equation}
 \dd s^2=\dd r^2+F(r)^2\dd\varphi^2,\qquad
 F=R\sinh\rho,\quad \rho=r/R,\quad
 A_\varphi=b(\cosh\rho-1),\quad b=BR^2 .
 \label{eq:hyperbolic-polar}
\end{equation}
Here $R>0$, $r>0$, and $B=A_\varphi'/F$ is the constant magnetic
field. In the rotating spin frame set $J=-i\hbar\partial_\varphi$,
$\delta_r=\partial_r+F'/(2F)$ and $\Gamma=\sigma_3$. The conserved
first-order operators are
\begin{align}
 Q_+&=\sigma_+\left[\hbar\delta_r+\frac{J-A_\varphi}{F}\right],
 \qquad Q_-=Q_+^\dagger,\qquad \mathcal H_P=\tfrac12\{Q_+,Q_-\},
 \nonumber\\
 R_-&=e^{-i\varphi}\left[\hbar\partial_r+\frac{F'}FJ
                    +\frac{A_\varphi}{F}-\frac{\hbar\Gamma}{2F}\right].
 \label{eq:hyperbolic-parent}
\end{align}
As on the plane, $R_-$ commutes with $Q_\pm$ and $\mathcal H_P$
and lowers $J$ by $\hbar$. Hence $Q_+R_-$ preserves
$K=J+\hbar\Gamma/2$. The same three-channel route
\eqref{eq:pauli-route}, now with radial measure $F\dd r$, gives
\begin{equation}
 q_-q_+=h_-(h_-+2\beta_R),\qquad
 \beta_R=\frac{k(k+2BR^2)}{4R^2}.
 \label{eq:hyperbolic-closure}
\end{equation}
Here $h_\pm$ are the two reduced Pauli blocks, $q_+$ is half the
reduced product, and $\beta_R$ is a signed energy shift.
As in Eq.~\eqref{eq:pauli-shift-conservation}, this shift is sectorwise:
adding $\beta_R(K)$ to the parent would preserve the composite but
would generally change the commutators of its first-order factors.
With $H_\pm=h_\pm+a+\beta_R$ and $k\ne0$, their
second-order SUSY data are
\begin{equation}
 W=\frac{k}{2R}\coth\frac{r}{2R},\qquad
 C=-\beta_R^2,\qquad
 \Delta_W=\frac{\hbar^2(2\cosh\rho+1)}
                         {4R^2\sinh^2\rho}.
 \label{eq:hyperbolic-target}
\end{equation}
Thus $\beta_R$ need not equal the positive root
$\sqrt{-C}$. At fixed $B$, the flat limit $R\to\infty$ gives
$W=k/r$, $\beta_R=Bk/2=\omega k$, and Eq.~\eqref{eq:pauli-target}.
At $B=0$ and finite $R$, however, $C=-k^4/(16R^4)$ is nonzero:
curvature enters the reduced invariant even without a magnetic field.
Conversely, $B=-k/(2R^2)$ gives $C=0$ at nonzero $k$.
For spinors extending through the polar axis, $k=\hbar\ell$ with
$\ell\in\mathbb Z$. At fixed $B,R,\hbar$, this additional sector
exists precisely when $-2BR^2/\hbar\in\mathbb Z\setminus\{0\}$;
the unreduced model allows arbitrary uniform $B$. Thus $C=0$ need not
imply zero cyclic momentum outside the Witten relation
$C=-\varepsilon\Omega^2k^2$ with $\Omega\ne0$.
The correction $\Delta_W$ varies with $r$ on this constant-curvature
surface, again distinguishing it from a scalar-curvature coupling.
Appendix~\ref{app:hyperbolic-polar} gives the two factors and their
closure, including the $k=0$ specialization
$q_+=-(H_\pm-a)$: even at nonzero curvature and field, the scalar
charge then reduces to a shifted Hamiltonian. As in the planar
$k=0$ sector, the candidate singlet at $E=a$ is excluded by its
behavior at the polar axis.

For a regular full-line problem, we now reduce along a different
Killing field. Geodesic coordinates adapt the reduction to translation
in a noncompact coordinate $y$, rather than rotation about the polar
axis. On the same hyperbolic plane, take
\begin{equation}
 \dd s^2=\dd x^2+R^2\cosh^2(x/R)\dd y^2,\qquad
 A_y=BR^2\sinh(x/R),\qquad x,y\in\mathbb R .
 \label{eq:hyperbolic-geodesic}
\end{equation}
The coordinate $y$ is dimensionless. Removing the measure and fixing
$-i\hbar\partial_y=j$ reduces the first-order charge to
\begin{equation}
 A_j=\hbar\partial_x+\frac jR\operatorname{sech}(x/R)
                         -BR\tanh(x/R),\qquad
 h_-(j)=\tfrac12A_j^\dagger A_j,\quad
 h_+(j)=\tfrac12A_jA_j^\dagger .
 \label{eq:geodesic-A}
\end{equation}
Choose $B=-\hbar/R^2$ and $j=0$, and put
$\gamma=\hbar/R$, $\epsilon=\gamma^2/2$, $p=-i\hbar\partial_x$.
Then
\begin{equation}
 A_0=\hbar\partial_x+\gamma\tanh(x/R),\qquad
 h_+=\frac{p^2}{2}+\epsilon,\qquad
 h_-=\frac{p^2}{2}+\epsilon-\gamma^2\operatorname{sech}^2(x/R).
 \label{eq:regular-PT-pair}
\end{equation}
After a common subtraction of $\epsilon$, these are a free particle
and the reflectionless P\"oschl--Teller system with one bound state.
Because $p$ commutes with $h_+$, the product $pA_0$ is a second-order
intertwiner in addition to $A_0$. In the normalization of
Section~\ref{sec:target}, take $q_+=ipA_0/2$. Keeping the unshifted
$h_\pm$ as $H_\pm$ gives
\begin{equation}
 W=\frac{\hbar}{2R}\tanh(x/R),\qquad
 U=\frac{\hbar^2}{4R^2}[1-3\operatorname{sech}^2(x/R)],\qquad
 a=\frac{\epsilon}{2},\quad C=-\frac{\epsilon^2}{4}.
 \label{eq:regular-PT-data}
\end{equation}
Thus $q_-q_+=h_-(h_--\epsilon)$. Although $W$ vanishes at $x=0$,
the complete $U$ is smooth: its two separately singular terms
$\Delta_W-C/W^2$ cancel as described in Appendix~\ref{app:zeros}.
The bound state proportional to $\operatorname{sech}(x/R)$ has
energy $0=a-\sqrt{-C}$ and is a physical singlet. The other root,
$\epsilon$, is a continuum threshold with no normalizable singlet.
This realizes the lower root without contradicting
Eq.~\eqref{eq:parent-energy-bound}, which was derived for the
centrally extended Witten parent.

Together, $A_0$ and $pA_0$ give the known exotic $N=4$ nonlinear
supersymmetry: two first-order and two second-order Hermitian charges,
whose mixed anticommutators give a third-order Lax--Novikov integral
\cite{CorreaJakubskyNietoPlyushchay2008,PlyushchayNieto2010}.
Appendix~\ref{app:hyperbolic-geodesic} gives the four matrix charges
and their algebra, and shows why this additional product intertwines
only at $j=0$. The first-order charge is conserved on the full surface;
the extra second-order symmetry belongs to the selected Fourier fiber,
where the upper partner becomes free. The bound state is normalizable
in $x$, not on the noncompact surface. Moreover, the tuning
$B=-\hbar/R^2$ defines a quantum-dependent family: at fixed $R$,
its reflectionless depth $\gamma^2$ and
$C=-\hbar^4/(16R^4)$ tend to zero with $\hbar$. It is not a
classical limit at fixed nonzero magnetic field.

The same free-particle--one-bound-state P\"oschl--Teller pair also
occurs in the AdS$_2$ Aharonov--Bohm reduction of
Ref.~\cite{CorreaJakubskyPlyushchay2009}. At half-integer flux, reduced
isometry ladders give Darboux operators; their products generate
higher-order intertwiners and Lax--Novikov integrals. For the pair
considered here, the first- and second-order intertwiners agree with
$A_0$ and $pA_0$, up to scales and phases. The parent systems differ:
that construction uses a spinless particle on Lorentzian AdS$_2$ with
an Aharonov--Bohm connection, whereas the present one uses a spin-$1/2$
particle on the Riemannian hyperbolic plane in a uniform magnetic field.

\section{Scalar and matrix factorizations}
\label{sec:limits}

The Witten construction uses an indefinite surface for $C>0$.
To determine which part of this conclusion depends on that particular
construction, we compare scalar and matrix intermediates, then give a
first-order surface example with constant $W$ and a positive inner
product. Positivity here refers to the state-space inner product,
not to the eigenvalues of the potential matrix.

\subsection{Scalar intermediates and the earlier factorization}

Successive Darboux transformations construct higher-order intertwiners
through intermediate scalar Hamiltonians \cite{Crum1955}.
Polynomial supersymmetry also includes irreducible transformations,
for which no regular real scalar first-order chain exists
\cite{AndrianovIoffeNishnianidze1995,AndrianovIoffe2012}.
Suppose first-order scalar operators $B_1,B_2$ connect the partners
through a scalar Hermitian $H_I$, with real factorization energies
$e_0,e_2$:
\begin{equation}
 \begin{aligned}
 H_-&=\tfrac12B_1^\dagger B_1+e_0,&
 H_I&=\tfrac12B_1B_1^\dagger+e_0
     =\tfrac12B_2^\dagger B_2+e_2,\\
 H_+&=\tfrac12B_2B_2^\dagger+e_2,&
 q_+&=\tfrac12B_2B_1.
 \end{aligned}
 \label{eq:real-chain}
\end{equation}
Using $H_IB_1=B_1H_-$ in the product of the composite and its
adjoint gives
$q_-q_+=(H_--e_0)(H_--e_2)$. Comparison with the second-order
SUSY polynomial therefore yields
\begin{equation}
 a=\frac{e_0+e_2}{2},\qquad
 C=-\frac{(e_0-e_2)^2}{4}\leq0.
 \label{eq:real-obstruction}
\end{equation}
The restriction concerns a \emph{real scalar} intermediate with these
adjoint factorizations. It does not exclude a complex scalar
intermediate or a matrix intermediate.

The distinction is explicit in the factorization of
Ref.~\cite{Plyushchay2017}, Eqs.~(2.25)--(2.29). In our notation,
on an interval on which $W$ does not vanish, choose either root $\beta$ of
$\beta^2=-C$ and put
\begin{equation}
 g=\frac{\hbar W'}{2W},\qquad
 B_{1,\beta}=\hbar\partial_x+W-g+\frac\beta W,\qquad
 B_{2,\beta}=\hbar\partial_x+W+g-\frac\beta W.
 \label{eq:beta-factors}
\end{equation}
The superscript $t$ denotes formal transposition: it reverses $\partial_x$ without
complex conjugating coefficients. Direct multiplication gives
\begin{align}
 2q_+&=B_{2,\beta}B_{1,\beta},\qquad 
 2q_-=B_{1,\beta}^{t}B_{2,\beta}^{t},\nonumber\\
 H_-&=\tfrac12B_{1,\beta}^{t}B_{1,\beta}+a-\beta,
 \qquad H_+=\tfrac12B_{2,\beta}B_{2,\beta}^{t}+a+\beta,\nonumber\\
 H_I(\beta)&=\tfrac12B_{1,\beta}B_{1,\beta}^{t}+a-\beta
 =\tfrac12B_{2,\beta}^{t}B_{2,\beta}+a+\beta.
 \label{eq:beta-chain}
\end{align}
These identities imply
$H_I(\beta)B_{1,\beta}=B_{1,\beta}H_-$ and
$H_+B_{2,\beta}=B_{2,\beta}H_I(\beta)$.
For $C<0$, $\beta$ is real, so formal transposition agrees with
the adjoint, recovering
Eq.~\eqref{eq:real-chain}. For $C=0$ the two factorization energies
coincide. For $C>0$, $\beta$ is imaginary, these operations differ, and the
intermediate scalar potential is generally complex, although the
endpoint Hamiltonians and the full $q_\pm$ retain their usual adjoint
relations.
The complex roots give one obstruction to a regular real scalar
chain. Conversely, real or coincident roots do not guarantee global
reducibility: an intermediate potential can be singular even when
both endpoint potentials are regular \cite{AndrianovIoffe2012}.
The displayed factors specify a local chain on a regular interval.

The surface construction explains where these scalar intermediates
come from. After cyclic reduction and removal of the radial measure,
its one-fermion Hamiltonian, including the energy shift $a$, is
\begin{equation}
 \begin{aligned}
 H_{I,W}&=H_0\id_2-i\hbar k\frac{F'}{F^2}\mathcal R_\varepsilon,
 \qquad \beta^2=\varepsilon\Omega^2k^2=-C,\\
 H_0&=a+\frac12\left[-\hbar^2\partial_x^2+W^2+g^2-\hbar g'
                           +\frac{\beta^2}{W^2}\right].
 \end{aligned}
 \label{eq:surface-intermediate}
\end{equation}
The two constant eigenvectors of $\mathcal R_\varepsilon$ diagonalize
this matrix into
\begin{equation}
 H_I(\beta)=H_0-\hbar\beta\frac{W'}{W^2},\qquad
 H_I(-\beta)=H_0+\hbar\beta\frac{W'}{W^2}.
\end{equation}
These are exactly the scalar intermediates of Eq.~\eqref{eq:beta-chain}.
For the Riemannian surface, $\beta$ is real and the polarization channels
are orthogonal in a positive inner product. For the Lorentzian surface,
$\beta$ is imaginary: the channels have complex-conjugate potentials,
and their polarization vectors are null and paired with each other by
the indefinite inner product. An individual channel is then generally
non-Hermitian, while the full two-component Hamiltonian is formally
self-adjoint in the indefinite pairing. This identifies the relation
between the indefinite fermion metric and the complex scalar
factorization. Appendix~\ref{app:middle-channels} derives the matrix and
the scalar factors by resolving its two polarizations.

\subsection{Local factorization through a two-component Hamiltonian}

Matrix supersymmetry has its own hidden differential symmetries and
reducibility questions \cite{AndrianovEtAl1997Matrix}.
Here the endpoint wavefunctions remain scalar while the intermediate
has two components. The first-order maps are consequently rectangular:
a column $S$ and a row $T$,
\begin{equation}
 H_-\ \xrightarrow{S}\ H_I\ \xrightarrow{T}\ H_+,
 \qquad S:\mathbb C\to\mathbb C^2,
 \quad T:\mathbb C^2\to\mathbb C,
 \qquad TS=2q_+.
 \label{eq:rectangular-chain}
\end{equation}
The arrows mean intertwining, $H_IS=SH_-$ and $H_+T=TH_I$:
each step maps an energy solution to another solution at the same
energy, or to zero. The spaces $\mathbb C$ and $\mathbb C^2$
specify the component dimensions; both maps act on functions of $x$.

Here is a local construction for every smooth compatible pair $W,U$.
Choose a constant momentum scale $\lambda>0$ and real functions
$z(x),\eta(x)$ satisfying
\begin{equation}
 \hbar z'=z^2+\eta^2-\lambda^2-U,\qquad
 \lambda \eta'=-W'z-Wz'-\tfrac\hbar2 W''.
 \label{eq:matrix-ode}
\end{equation}
A prime again denotes $\dd/\dd x$. Substituting the first equation
into the second makes this an ordinary first-order initial-value
system for $z,\eta$. Define
\begin{equation}
 S=\begin{pmatrix}\hbar\partial_x+W+z\\\lambda+\eta\end{pmatrix},\qquad
 T=\begin{pmatrix}\hbar\partial_x+W-z&\lambda-\eta\end{pmatrix}.
 \label{eq:ST}
\end{equation}
The intermediate Hamiltonian is
\begin{equation}
 H_I=-\frac{\hbar^2}{2}\partial_x^2\id_2+
 \begin{pmatrix}
 V_0+\hbar z'&\hbar \eta'\\
 \hbar \eta'&V_0-\dfrac\hbar\lambda(W\eta)'
 \end{pmatrix},\qquad V_0=a+\frac{W^2+U}{2}.
 \label{eq:matrix-HI}
\end{equation}
Its real symmetric potential makes $H_I$ formally self-adjoint in
$L^2(\dd x)\otimes\mathbb C^2$ with its usual positive inner product.
The first equation in \eqref{eq:matrix-ode} ensures $TS=2q_+$.
The remaining intertwining conditions reduce to
$\Bol[W]=0$, already required by the target. Appendix~\ref{app:matrix}
shows these cancellations explicitly.

For smooth $W,U$ and $\hbar,\lambda>0$, the first-order system
\eqref{eq:matrix-ode} has smooth coefficients and hence a regular local
solution for any finite initial data. It contains no division by $W$,
so the construction applies for every sign of $C$ and also through
admissible zeros of $W$, as specified in Section~\ref{sec:target}.
Its scope is local: the solutions $z,\eta$ may develop poles outside
the initial interval. The scale $\lambda$ and the initial values of
$z,\eta$ specify a family of chains, with the $x$-independent first
integral
\begin{equation}
 e=Wz+\tfrac\hbar2 W'+\lambda\eta.
 \label{eq:matrix-chain-constant}
\end{equation}
This is one parameter of the chain, not an additional dynamical
integral of the target system. Global continuation and self-adjoint
realizations require separate analysis. Appendix~\ref{app:matrix}
gives necessary spectral bounds for any such realization.
The scalar obstruction does not apply because
$q_-q_+=S^\dagger(T^\dagger T)S/4$ contains a \emph{matrix}
operator $T^\dagger T$ on the intermediate space. It cannot be replaced
by a scalar shifted copy of $H_I$ as in Eq.~\eqref{eq:real-chain}.

For an explicit nonconstant $C>0$ target, take dimensionless units
$\hbar=1$, $a=0$, $W=1+x^2$ and $U=0$. Equation~\eqref{eq:I}
gives $C=1$, and
\begin{equation}
 H_\pm=\tfrac12[-\partial_x^2+(1+x^2)^2\pm4x].
 \label{eq:positive-target}
\end{equation}
Choose $\lambda=1$, $z(0)=\eta(0)=0$. The local solution begins as
$z=-x+x^3/3+O(x^5)$ and $\eta=2x^3/3+O(x^5)$.
The potential matrix in Eq.~\eqref{eq:matrix-HI} is regular, with
value $\diag(-1/2,1/2)$ at the origin. This gives a regular local
formally Hermitian intermediate for a nonconstant $C>0$ target, with both
first-order maps satisfying the intertwining identities.
Here $W,U$ are even: spatial parity $\mathsf P\psi(x)=\psi(-x)$
also gives $\mathsf P H_-\mathsf P=H_+$ and
$\mathsf P q_+\mathsf P=q_-$. This explains the spectral equivalence
of these full-line partners. The matrix-chain construction supplies
the first-order maps and intermediate Hamiltonian for general
compatible $W,U$, without assuming parity.

The dimensions $1\to2\to1$ match Witten's fermion-degree sectors,
but the conditions differ: Eq.~\eqref{eq:matrix-ode} enforces two
intertwinings and their product, whereas the surface construction also
requires nilpotent odd generators on all four components and the full
$N=4$ algebra. A local matrix chain therefore does not by itself
identify $H_I$ with the Witten intermediate or supply a complete
first-order surface supersymmetry.

\subsection{A constant-superpotential parent with a positive inner product}

There is also an exact surface construction with a positive-definite
state-space inner product for $C>0$ when
$W=w\ne0$ is constant. It is useful because it rules out a universal
exclusion based on the sign of $C$ alone. Put
$\mu=\sqrt C/|w|>0$, so $U=-\mu^2$, and use length coordinates
$x,y$ on a flat cylinder, with $y$ periodic. Introduce
\begin{equation}
 X=\hbar\partial_x+w,\qquad Y=\hbar\partial_y-\mu,
 \qquad
 D_1=\epsilon_1X+\epsilon_2Y,\qquad
 D_2=\epsilon_2X-\epsilon_1Y.
 \label{eq:constant-D}
\end{equation}
The operators $\epsilon_1,\epsilon_2$ create the two fermions, or
equivalently wedge with the two orthonormal one-forms. In form notation,
\begin{equation}
 D_1=\hbar\dd+\alpha\wedge,\qquad
 \alpha=w\dd x-\mu\dd y,\qquad \dd\alpha=0.
 \label{eq:constant-closed-form}
\end{equation}
If the period of $y$ is $L_y$, then
$\oint\alpha=-\mu L_y\ne0$: this closed one-form is not globally
exact on the cylinder. In contrast to the deformation $\dd\chi=W\dd x$
used earlier, it has a real component along the cyclic direction.
That component remains $Y=-\mu$ in the zero Fourier mode and is
distinct from the imaginary term supplied by a real cyclic momentum.
The constant coefficients and fermionic anticommutation relations make
$D_1,D_2$ nilpotent and mutually anticommuting. Both commute with
\begin{equation}
 \mathcal G=a+\tfrac12[-\hbar^2(\partial_x^2+\partial_y^2)
                                  +w^2-\mu^2]\id_4.
 \label{eq:constant-G}
\end{equation}
The diagonal anticommutators are
$\{D_i,D_i^\dagger\}=2(\mathcal G-a)+2\mu^2$. The mixed ones are
\begin{equation}
 \{D_1,D_2^\dagger\}=Z,\qquad \{D_2,D_1^\dagger\}=-Z,\qquad
 Z=2\hbar(\mu\partial_x+w\partial_y).
\end{equation}
The translation $Z$ is skew-adjoint. Replacing $D_2$ by $iD_2$
makes both mixed anticommutators equal to the Hermitian central
operator $-iZ$. Thus the algebra closes linearly on $\mathcal G$,
a translation and the identity. Appendix~\ref{app:constant} displays
the four Hermitian first-order generators, derives their brackets,
and gives a common invariant core for their global realization on the
complete cylinder.

The product from zero-forms to two-forms has coefficient $X^2+Y^2$.
In the zero Fourier mode, $\partial_y=0$, it becomes
\begin{equation}
 \tfrac12(X^2+Y^2)=\tfrac12[(\hbar\partial_x+w)^2+\mu^2]=q_+.
\end{equation}
The endpoint Hamiltonians are identical and equal to $H_\pm$,
giving $C=w^2\mu^2>0$ with a positive inner product. Non-exactness
is not essential to this local algebra: on the plane the same one-form
is $\dd(wx-\mu y)$. The cylinder makes the zero Fourier mode
normalizable in the cyclic direction. The essential change from the
earlier model is the real transverse deformation, rather than simply
a different choice of real cyclic momentum.

One can test a more general extension by allowing both $W$ and the
cylinder radius to vary. Retain the length coordinates $x,y$ and take
$\dd s^2=\dd x^2+f(x)^2\dd y^2$, with dimensionless $f>0$, and
$\alpha=W(x)\dd x-\mu\dd y$, with constant $\mu\ne0$.
The operators $d_\alpha=\hbar\dd+\alpha\wedge$ and
$d_\alpha^c=\mathcal R d_\alpha\mathcal R^{-1}$ remain nilpotent
and anticommute. Their positive-adjoint Hamiltonians agree on zero-
and two-forms, but their one-form blocks differ by
\begin{equation}
 \mathcal H_c^{(1)}-\mathcal H_\alpha^{(1)}
 =\hbar\begin{pmatrix}
 -s_f&-2\mu f'/f^2\\
 -2\mu f'/f^2&s_f
 \end{pmatrix},\qquad s_f=W'-\frac{f'}fW,
 \label{eq:constant-extension-defect}
\end{equation}
where $\mathcal H_\alpha=\{d_\alpha,d_\alpha^\dagger\}/2$ and
$\mathcal H_c=\{d_\alpha^c,(d_\alpha^c)^\dagger\}/2$.
For $\mu\ne0$, equality forces $f'=0$ and then $W'=0$.
At $f=1$, the full difference reduces to
$\hbar W'\diag(0,-1,1,0)$.
Appendix~\ref{app:constant} derives the mismatch. Thus varying the
radius does not remove the obstruction within this closed-one-form
construction with its standard rotated companion. A positive
first-order realization for arbitrary nonconstant $W$ and $C>0$
would require going beyond this ansatz. The nonlinear Pauli example
of Section~\ref{sec:pauli-extensions}, which starts with second-order
charges, leaves that problem open.

\section{Conclusions and outlook}
\label{sec:conclusions}

The Witten construction derives the complete second-order system from
two conserved first-order raising operators in a centrally extended
$N=4$ surface model. Their product connects the scalar zero- and
two-fermion sectors through a two-component intermediate. The condition
$W=\Omega F$ ensures their conservation and identifies the radial-measure
term with $\Delta_W$, supplying the factor $|W|^{-1/2}$ selected by
the anomaly-free prescription of Ref.~\cite{Plyushchay2017}.

Isothermal coordinates relate this construction to the auxiliary
projective geometry: the Hill equation is the separated Laplace or
wave equation for the gauge-transformed charge kernel, and the
isothermal coordinate's Schwarzian gives $\Delta_W$. The selected
stabilizer generator is the conformal gradient of the Witten deformation.
The full auxiliary
$\mathfrak{sl}(2,\mathbb R)$ need not be a symmetry of the physical
Hamiltonian; the AdS$_2$ example makes this distinction explicit.

Within this construction, the cyclic term $C=-\varepsilon\Omega^2k^2$
gives $C\leq0$ for Riemannian parents and $C\geq0$ for Lorentzian
parents. Their one-fermion blocks resolve into real scalar channels
for $C\leq0$ and generally complex ones for $C>0$.
The positive reduced theory in negative signature requires a separate
adjoint and pairing; it is not a unitary image of the indefinite parent.

Pauli reduction instead combines an odd first-order charge with an
even spatial symmetry. Its flat example shows that $\Delta_W$ depends
on measure, spin transport and angular shifts, not curvature alone.
On the hyperbolic plane, a conserved polar composite gives
$C=-[Bk/2+k^2/(4R^2)]^2$, whereas the geodesic
free-particle--P\"oschl--Teller pair gains its additional charge in a
selected Fourier sector. Nonlinear Pauli models also realize $C>0$
on the Euclidean plane with a positive-definite Hilbert-space inner
product, but already have second-order parent charges.

A positive two-component intermediate permits local first-order chains
for arbitrary $C$, beyond the real scalar-chain restriction. Such a
chain does not automatically supply the full surface superalgebra.
Constant $W$ admits a global positive first-order realization on a flat
cylinder with a real transverse deformation. Within its closed-one-form
ansatz, allowing a variable radius still forces $W$ to remain constant.
A positive first-order surface origin for arbitrary nonconstant $W$
and $C>0$ remains open.

The real roots of the quadratic supersymmetry polynomial give candidate
singlet energies, reproduced by the auxiliary Hill symmetry representation
on local charge kernels. Normalizability and compatible operator domains
determine which candidates are physical singlets. The Dirichlet radial
oscillator excludes the lower full-line oscillator singlet, while the
regular Pauli pair realizes the lower root excluded by the positive
Witten bound. At admissible simple zeros $x_0$, the quantum correction
removes the classical repulsive potential barrier and makes both
Hamiltonians and charges regular, with $C=-\hbar^2W'(x_0)^2/4$.
Physical continuation still requires compatible domains.

The next question is which higher-order families descend from conserved
first-order composites, beyond known special reductions
\cite{AnabalonPlyushchay2003}. The third-order intertwining relations of
Ref.~\cite{IoffeNishnianidze2004} provide a concrete target.
Higgs--Hahn and Racah reductions, Killing--Yano symmetries and
Eisenhart--Duval lifts offer further comparisons
\cite{FrappatEtAl2019,GaboriaudEtAl2018,Cariglia2014,
DuvalGibbonsHorvathy1991}. Their parent integrals need not be first
order, and Dirac symmetries descend only when they preserve the
selected subspace \cite{Cariglia2012}.

\medskip
\noindent\textbf{Acknowledgements.}
The work was partially supported by the FONDECYT Project 1242046. ChatGPT by OpenAI was used as an aid in editing and checking analytical computations. The authors take responsibility for the final content.

\appendix

\section{Deriving the intertwiner and continuing it through zeros}
\label{app:intertwiner}

\noindent\textbf{The relative sign of the scalar term.}
Keep $H_\pm$ in Eq.~\eqref{eq:H} and start with a real second-order
charge. The coefficient of $\partial_x^3$ in the intertwining defect
$H_+\widehat q-\widehat qH_-$ forces the leading coefficient of
$\widehat q$ to be constant. After normalization, write
\begin{equation}
 \widehat q=\tfrac12(\hbar^2\partial_x^2+2\hbar f\partial_x+g_0).
 \label{eq:general-q}
\end{equation}
The coefficient of $\partial_x^2$ in the same defect is
$\hbar^3(W'-f')$, so $f=W+b$ with constant $b$.
The reparametrization $W\mapsto W+b$, $U\mapsto U-2bW-b^2$
preserves $H_\pm$ and allows us to set $f=W$.
The coefficient of $\partial_x$ in $H_+\widehat q-\widehat qH_-$ is then
$\hbar^2(\hbar W''+2WW'-U'-g_0')/2$. Its vanishing gives
$g_0=\hbar W'+W^2-U+c$.
Absorbing $c$ by $U\mapsto U-c$, $a\mapsto a+c/2$ again preserves
$H_\pm$ and yields exactly the charge~\eqref{eq:q}.
Thus the derivative coefficients fix the relative sign $-U$ before
imposing the remaining scalar condition. The latter is
Eq.~\eqref{eq:defect}; its first integral and solution are already
Eqs.~\eqref{eq:I} and \eqref{eq:correction}.

\smallskip
\noindent\appendixlabel{app:zeros}\textbf{Continuation through zeros.}
For smooth $U$, the Bol equation is a regular linear third-order
ODE. Uniqueness excludes triple or higher zeros of a nontrivial
solution, including infinitely flat zeros.
At a simple zero $x_0$, its first integral requires
$C=-\hbar^2W'(x_0)^2/4<0$; at a double zero it requires $C=0$,
and the Bol equation also requires $W'''(x_0)=0$.
Conversely, for a prescribed smooth $W$, these conditions suffice
to extend $U$ smoothly through the zero.
Indeed, write
\begin{equation}
 U=\frac{\mathcal N_W}{W^2},\qquad
 \mathcal N_W=\frac{\hbar^2}{2}WW''-\frac{\hbar^2}{4}W'^2-C,
 \qquad \mathcal N_W'=\frac{\hbar^2}{2}WW'''.
 \label{eq:zero-numerator}
\end{equation}
For an admissible simple zero, $\mathcal N_W(x_0)=0$ and
$\mathcal N_W'=O(x-x_0)$, so numerator and denominator have a common
smooth factor $(x-x_0)^2$. At an admissible double zero,
$\mathcal N_W'=O((x-x_0)^3)$ and the common factor is $(x-x_0)^4$.
After cancellation, the remaining denominator is nonzero. The limits are
\begin{equation}
 U(x_0)=
 \begin{cases}
 \displaystyle\frac{\hbar^2W'''(x_0)}{4W'(x_0)},&\text{simple zero},\\[6pt]
 \displaystyle\frac{\hbar^2W''''(x_0)}{4W''(x_0)},&\text{admissible double zero}.
 \end{cases}
 \label{eq:zero-values}
\end{equation}
After one application of l'H\^{o}pital's rule, cancellation of the
common factor $W$ leaves $\hbar^2W'''/(4W')$.
This gives the simple-zero limit directly; at an admissible double
zero one further application gives the second ratio.

On a connected interval with smooth $U$, the zeros of a nontrivial
$W$ cannot accumulate in the interior: Rolle's theorem would give
$W=W'=W''=0$ at an accumulation point, and ODE uniqueness would
force $W\equiv0$. Since $C$ is fixed, any zeros are all simple with
$|W'|=2\sqrt{-C}/\hbar$ when $C<0$, or all admissible doubles
when $C=0$; for $C>0$ there are none. Simple and double zeros
therefore cannot coexist in one regular system. Neither $C<0$ nor
$C=0$ requires zeros to occur. These statements concern continuation
through zeros, not SUSY on punctured intervals.

\section{Classical normal forms and their quantum limits}
\label{app:classical}

Use the classical variables and charges of
Section~\ref{sec:target}, and write
$h_0=H_{\rm cl,body}-a$, so that $H_{\rm cl}=a+h_0+2W'N$ and
$|B|^2=h_0^2+C$. Charge zeros mean zeros of the ordinary coefficient
$B$, not the result of setting the Grassmann variables to zero.
The transformations below are local on a regular phase-space patch.

\smallskip
\noindent\textbf{Canonical phase change.}
For real $G(x,p)$, the canonical flow generated by $GN$ gives
\begin{equation}
 \Theta^\pm=e^{\pm iG}\theta^\pm,\qquad
 X=x+G_pN,\qquad P=p-G_xN,\qquad
 \mathcal N=\Theta^+\Theta^-=N.
 \label{eq:classical-map}
\end{equation}
The bosonic series stops because $N^2=0$, whereas the fermionic series
exponentiates. The shifts of $x,p$ ensure canonicity. Writing
$\nu=2W'$, substitution of the inverse shifts into
$H_{\rm cl}=a+h_0+\nu N$ yields
\begin{equation}
 H_{\rm cl}=a+h_0(X,P)+[\nu+\{G,h_0\}](X,P)\mathcal N,
 \qquad Q_+=e^{-iG(X,P)}B(X,P)\Theta^+.
 \label{eq:classical-coupling-change}
\end{equation}
There are no further terms in the charge because
$\mathcal N\Theta^\pm=0$. Thus removing a phase from $B$ also changes
the fermion coupling by $\{G,h_0\}$.

\smallskip
\noindent\textbf{The three signs of $C$.}
For $C=-\beta^2<0$, set $w_\pm=W\pm\beta/W$, $\beta>0$.
The factorization $2B=(w_-+ip)(w_++ip)$ has a nonvanishing second
factor wherever real $W\ne0$. Choose $G=\arg(w_++ip)$.
Since $h_0=\beta+(p^2+w_-^2)/2$ and $w_+w_+'=w_-w_-'$,
\[
 \{G,h_0\}
 =-\frac{p^2w_+'+w_+w_-w_-'}{p^2+w_+^2}=-w_+'.
\]
Consequently $2W'+\{G,h_0\}=w_-'$, and the transformed system is
\begin{align}
 H_{\rm cl}&=a+\beta+H_1,\qquad
 H_1=\tfrac12[P^2+w_-(X)^2]+w_-'(X)\mathcal N,\nonumber\\
 s_\pm&=\frac{w_-(X)\pm iP}{\sqrt2}\Theta^\pm
       =\frac{Q_\pm}{\sqrt{H_{\rm cl}-a+\beta}},
 \qquad \{s_+,s_-\}=-iH_1.
 \label{eq:classical-linear-normalization}
\end{align}
The additional normalization is regular because the denominator has
strictly positive bosonic part. The original charges still obey the
quadratic algebra; the normalized $s_\pm$ obey linear supersymmetry.
Their zeros, if present, occur at $p=0$, $W^2=\beta$, and reach only
the upper root, since $h_0\geq\beta$.

For $C=0$, the holomorphic coefficient $B=(W+ip)^2/2$ has winding
number of absolute value two around a simple zero $(x_0,0)$ of
$W+ip$. Its phase can be removed away from that point, but the
rotation does not extend regularly through it.
For $C>0$, instead, $|B|^2=h_0^2+C>0$ everywhere on the patch.
Choosing $G=\arg B$, charge conservation gives
$\{G,h_0\}=-2W'$. Equation~\eqref{eq:classical-coupling-change}
then removes the fermionic term, and
\begin{equation}
 \frac{Q_\pm}{\sqrt{(H_{\rm cl}-a)^2+C}}=\Theta^\pm,
 \qquad \{\Theta^+,\Theta^-\}=-i.
\end{equation}
This is the $n=0$ normal form with conserved spectator fermions.
It leaves the bosonic potential, including its possible singularities,
unchanged.

\smallskip
\noindent\appendixlabel{app:quantum-domains}\textbf{Quantum domains and normalization.}
The spectral statements in Section~\ref{sec:target} require more than
formal differential identities. Let $\mathbb H$ be self-adjoint and
suppose that mutually adjoint charges satisfy
\begin{equation}
 \|\mathbb Q_+\Psi\|^2+\|\mathbb Q_-\Psi\|^2
 =\|(\mathbb H-a)\Psi\|^2+C\|\Psi\|^2
 \label{eq:closure-form}
\end{equation}
on a common charge domain $\mathcal D_0\subset D(\mathbb H)$ dense
in the graph norm of $\mathbb H$. The right-hand side is continuous
in that norm, so its nonnegativity extends to $D(\mathbb H)$.
The spectral theorem gives $P(\mathbb H)\geq0$, excluding both discrete
and continuous spectrum in $(a-\beta,a+\beta)$ when $C=-\beta^2<0$.
Test functions supported away from an endpoint need not be a graph
core for a chosen self-adjoint boundary condition, so a formal identity
on those functions alone is insufficient.

The regular half-line oscillator of Section~\ref{sec:cone} gives a
simple example of the distinction between a common invariant core and
the full Hamiltonian domain. At $g=0$ and $b>0$, the finite span of
odd oscillator eigenfunctions is an invariant core for the Dirichlet
Hamiltonian and its radial ladders. However,
$f(r)=r^2e^{-r/L}$, $L>0$, belongs to the Hamiltonian domain while
$(q_+f)(0)=\hbar^2\ne0$, so $q_+f$ violates the Dirichlet condition.
The ladder identities hold on the invariant core without requiring
$q_+$ to preserve every vector in the Hamiltonian domain.

Canonical transformations and unitary conjugations preserve the algebra
of the original charges. The normal forms above also redefine those
charges by functions of the Hamiltonian. Quantum inverse square roots
are generally nonlocal and require spectral and domain control.
For $C>0$, assume self-adjoint $H_\pm$ and a closed densely defined
$q_+$ with the operator identities
$q_+^\dagger q_+=P(H_-)$ and $q_+q_+^\dagger=P(H_+)$.
Both right-hand sides are bounded below by $C>0$, so the polar factor
\begin{equation}
 V=q_+[(H_--a)^2+C]^{-1/2},\qquad V^\dagger V=VV^\dagger=\id.
 \label{eq:polar-unitary}
\end{equation}
is a unitary identification of the sectors, generally not a local
differential operator. If $q_+$ also intertwines the spectral
resolutions of $H_-$ and $H_+$, then $VH_-V^\dagger=H_+$.
This last condition is essential: the norm identities identify only
$P(H_-)$ and $P(H_+)$, and $P(E)$ does not distinguish energies
equidistant from $a$.

For $C<0$, the inverse square root in
Eq.~\eqref{eq:classical-linear-normalization} is undefined or singular
on parts of the possible quantum spectrum. In particular, the
denominator vanishes at the full-line oscillator singlet $E=a-\beta$,
which has no classical counterpart because $h_0\geq\beta$.
Thus the regular classical passage to linear supersymmetry need not
extend to the whole quantum Hilbert space.

\section{The auxiliary Hill symmetry and the charge kernels}
\label{app:projective}

\noindent\textbf{The quadratic representation of $W$.}
Use the Hill basis and Wronskian of Section~\ref{sec:target}.
On an interval where a Hill solution $\psi=c_1\psi_1+c_2\psi_2$
is nonvanishing, set $\lambda=2\psi'/\psi$.
The Hill equation gives $\mathcal U=-\lambda'-\lambda^2/2$, hence
\[
 \mathscr L=(\partial_x+\lambda/2)(\partial_x-\lambda/2),\qquad
 \Bol=(\partial_x+\lambda)\partial_x(\partial_x-\lambda).
\]
These are local differential factorizations: the factors can be singular
at zeros of $\psi$ even when the complete operators are regular.
Since \((\partial_x-\lambda/2)\psi=0\) and
\((\partial_x-\lambda)\psi^2=0\), every Hill solution squared
solves the Bol equation, including through zeros by smoothness.
Applying this to $\psi_1,\psi_2$ and $\psi_1+\psi_2$ shows that
$\psi_1^2,\psi_1\psi_2,\psi_2^2$ are Bol solutions. Their Wronskian
$2\omega_0^3\ne0$ proves that they span its solution space, giving
Eq.~\eqref{eq:projective-data}.
The derivatives of $W$ are $W'=2\boldsymbol\psi^tM\boldsymbol\psi'$ and
$W''=2\boldsymbol\psi'^tM\boldsymbol\psi'-\mathcal U W$. Hence
\begin{equation}
 I[W]=4\big[(\boldsymbol\psi^tM\boldsymbol\psi)
             (\boldsymbol\psi'^tM\boldsymbol\psi')
             -(\boldsymbol\psi^tM\boldsymbol\psi')^2\big]
       =4\omega_0^2\det M,\qquad
 C=\hbar^2\omega_0^2\det M.
\end{equation}
This determinant identity holds also at zeros of $W$.
On a patch with $\psi_2\ne0$, the projective coordinate
$f=\psi_1/\psi_2$ satisfies $f'=-\omega_0/\psi_2^2$ and
$\{f;x\}=\mathcal U$. Changing the Hill basis transforms $f$
by a M\"obius map and leaves its Schwarzian unchanged.
For $f'>0$, the transformed Hill density is
$\widetilde\psi=\sqrt{f'}\,\psi$, and the vector-field coefficient becomes
$\widetilde W=f'W=-\omega_0(m_{11}f^2+2m_{12}f+m_{22})$.
These are the linear and quadratic polynomial spaces used in the
Casimir discussion in Section~\ref{sec:target}.

\smallskip
\noindent\textbf{The stabilizer and its constant matrix.}
The transformation law~\eqref{eq:projective-law-main} follows from
\[
 (\partial_y^2+\widetilde{\mathcal U}/2)
 [\phi'^{-1/2}\xi\circ\phi]
 =\phi'^{3/2}(\mathscr L\xi)\circ\phi.
\]
The weight $-1/2$ cancels the first derivative of $\xi$.
For the local flow $\partial_\tau\phi_\tau=W\circ\phi_\tau$,
$\phi_0(y)=y$, its density action
$T_\tau\xi=\phi_\tau'^{-1/2}\xi\circ\phi_\tau$ has generator
$\mathfrak D_W=W\partial_x-W'/2$. More explicitly,
\begin{equation}
 \mathscr L\mathfrak D_W-(\mathfrak D_W+2W')\mathscr L
 =-\tfrac12\Bol[W].
 \label{eq:hill-invariance}
\end{equation}
Thus $\Bol[W]=0$ preserves the Hill kernel. The two operators beside
$\mathscr L$ are Lie derivatives of weights $-1/2$ and $3/2$,
respectively; the identity expresses covariance between those density
spaces, rather than ordinary commutation.

For $W=m_{11}\psi_1^2+2m_{12}\psi_1\psi_2+m_{22}\psi_2^2$,
expand $W\psi_i'-W'\psi_i/2$. The $m_{11}$ terms cancel for $i=1$,
and the $m_{22}$ terms cancel for $i=2$. Factoring the remaining
Wronskians gives
\begin{align*}
 \mathfrak D_W\psi_1
 &=(m_{12}\psi_1+m_{22}\psi_2)(\psi_2\psi_1'-\psi_1\psi_2')
   =-\omega_0(m_{12}\psi_1+m_{22}\psi_2),\\
 \mathfrak D_W\psi_2
 &=(m_{11}\psi_1+m_{12}\psi_2)(\psi_1\psi_2'-\psi_1'\psi_2)
   =\omega_0(m_{11}\psi_1+m_{12}\psi_2).
\end{align*}
The resulting coefficient columns are precisely the columns of
$A=\omega_0MJ$ in Eq.~\eqref{eq:A}. In contrast, the basis-function
column obeys $\mathfrak D_W\boldsymbol\psi=A^t\boldsymbol\psi$.
This distinction fixes the order $MJ$ and explains why the matrix is
constant although $W$ generally depends on $x$.

Symmetry of $M$ gives $A^tJ+JA=0$. Writing
$T_\tau\xi=\boldsymbol\psi^tc(\tau)$ gives $\dot c=Ac$, so
$c(\tau)=e^{\tau A}c(0)$ preserves the pairing $\omega_0c^tJd$.
This is the stabilizer's action represented by Eq.~\eqref{eq:A}.
In the surface realization, $\psi_i=\sqrt F\,v_i(s)$ gives
$\operatorname{Wr}_x(\psi_1,\psi_2)=v_1v_{2,s}-v_{1,s}v_2$,
the conserved radial Green bracket of the harmonic equation.
It is neither the particle's phase-space form nor a positive
Hilbert inner product.

\smallskip
\noindent\textbf{Restriction to the charge kernels.}
With $\mathcal S'=W$, conjugation of the Hamiltonians on arbitrary
functions gives
\begin{align}
 e^{\mathcal S/\hbar}(H_--a)e^{-\mathcal S/\hbar}
 &=-\tfrac{\hbar^2}{2}\mathscr L+\hbar\mathfrak D_W,\nonumber\\
 e^{-\mathcal S/\hbar}(H_+-a)e^{\mathcal S/\hbar}
 &=-\tfrac{\hbar^2}{2}\mathscr L-\hbar\mathfrak D_W.
 \label{eq:gauge-full}
\end{align}
On a Hill solution, $-\hbar^2\xi''+U\xi=0$ removes the
$\mathscr L$ term, giving the kernel restrictions $a\id\pm\hbar A$.
On arbitrary functions the operators remain second order.

The formal eigenvalues $a\pm\sqrt{-C}$ become physical energies only
if the corresponding functions belong to the self-adjoint domains.
For $C>0$ they are complex and hence excluded. For $C<0$, the domain
and normalizability conditions decide which real roots occur.
For $C=0$ and $A\ne0$, the formal kernel contains a Jordan chain,
say $(H_--a)\Phi_0=0$ and $(H_--a)\Phi_1=\hbar\Phi_0$.
Both functions cannot belong to one self-adjoint domain: taking the
inner product with $\Phi_0$ would give
$0=\langle\Phi_0,(H_--a)\Phi_1\rangle=\hbar\|\Phi_0\|^2$
for $\Phi_0\ne0$.

\section{Surface matrices, composite algebra and scalar channels}
\label{app:witten-checks}
\label{app:surface-matrices}
\label{app:signed-matrices}

\noindent\textbf{Unreduced matrices for both signatures.}
The Riemannian and indefinite constructions can be displayed together.
Use $\dd s^2=\dd x^2+\varepsilon F^2\dd y^2$ and the component order
$(\Psi_0,\Psi_x,\Psi_y,\Psi_2)^t$, with fermion-degree block sizes
$1,2,1$. Setting $\varepsilon=1$ and $y=\varphi$ recovers the angular
construction. No cyclic momentum is fixed here. Put
$D=\hbar\partial_x$, $f=\hbar F'/F$, $X=D+W$, $T=X+f$ and
$Y=\hbar F^{-1}\partial_y$. In the measure $F\dd x\dd y$,
\begin{equation}
 D^\dagger=-D-f,\qquad X^\dagger=-D-f+W,\qquad
 T^\dagger=-D+W,\qquad Y^\dagger=-Y.
 \label{eq:surface-formal-adjoints}
\end{equation}
The positive componentwise adjoint is denoted by $\dagger$; the metric
adjoint is $A^\sharp=G_\varepsilon^{-1}A^\dagger G_\varepsilon$, with
$G_\varepsilon=\diag(1,1,\varepsilon,\varepsilon)$. They agree for
$\varepsilon=1$. The raising supercharges are
\begin{equation}
 Q=-i\begin{pmatrix}
 0&0&0&0\\X&0&0&0\\Y&0&0&0\\0&-Y&T&0
 \end{pmatrix},\qquad
 \widetilde Q=\begin{pmatrix}
 0&0&0&0\\-\varepsilon Y&0&0&0\\X&0&0&0\\0&-T&-\varepsilon Y&0
 \end{pmatrix}.
 \label{eq:surface-full-raising}
\end{equation}
Their lowering partners follow by taking the indicated metric adjoint;
for positive signature these are simply $Q^\dagger,\widetilde Q^\dagger$.
The relation \([D,Y]=-fY\) gives \(TY=YX\), which implies the nilpotency of both raising supercharges, \(Q^2=\widetilde Q^{\,2}=0\).
Writing $d=d_\chi=iQ$ and $c=d^c_{\chi,\varepsilon}=\widetilde Q$,
the products $dc$ and $cd$ have opposite $0\to2$ coefficients,
$TX+\varepsilon Y^2$ and $-TX-\varepsilon Y^2$, so the raising charges anticommute.

To show the complete Hamiltonian, define
\begin{equation}
 h_{{\rm b},\varepsilon}=-\frac{\hbar^2}{2}
 \left(\partial_x^2+\frac{F'}F\partial_x
                   +\frac{\varepsilon}{F^2}\partial_y^2\right)
 +\frac{W^2}{2},\qquad \mathfrak m=W'-\frac{F'}F W.
 \label{eq:surface-hb}
\end{equation}
Matrix multiplication gives
\begin{align}
 \mathcal H_\varepsilon
 &=\tfrac12\{Q,Q^\sharp\}
 =\diag(\mathcal H_\varepsilon^{(0)},
        \mathcal H_\varepsilon^{(1)},\mathcal H_\varepsilon^{(2)}),
 \label{eq:surface-H-blocks}\\
 \mathcal H_\varepsilon^{(0,2)}
 &=h_{{\rm b},\varepsilon}\mp\tfrac12(\hbar W'+fW),
 \nonumber\\
 \mathcal H_\varepsilon^{(1)}
 &=\left(h_{{\rm b},\varepsilon}-\tfrac\hbar2 f'\right)\id_2+
 \begin{pmatrix}\hbar\mathfrak m/2&\varepsilon fY\\-fY&-\hbar\mathfrak m/2\end{pmatrix}.
 \label{eq:surface-middle-expanded}
\end{align}
The middle block is $2\times2$, with
$fY=\hbar^2F'F^{-2}\partial_y$. For $W=\Omega F$, $\mathfrak m=0$
and $fW=\hbar W'$: the scalar blocks are
$h_{{\rm b},\varepsilon}\mp\hbar W'$ and the middle
diagonal entries coincide. The full Hamiltonian satisfies
$\mathcal H_\varepsilon^\sharp=\mathcal H_\varepsilon$.

\leavevmode\appendixlabel{app:witten-mixed}%
The companion bracket $\{\widetilde Q,\widetilde Q^\sharp\}/(2\varepsilon)$
has the same scalar blocks and replaces $\mathfrak m$ by $-\mathfrak m$ in the middle block.
Consequently the two Hamiltonians coincide precisely when $\mathfrak m=0$.
The mixed bracket $\{d,c^\sharp\}$ has endpoint coefficients
$2\varepsilon WY$ and middle block
\begin{equation}
 \{d,c^\sharp\}^{(1)}
 =\varepsilon\begin{pmatrix}2WY&2\hbar\mathfrak m\\0&2WY\end{pmatrix}.
\end{equation}
With $W=\Omega F$, this gives $\{d,c^\sharp\}=2i\varepsilon\Omega K$,
where $K=-i\hbar\partial_y$. Multiplication by $-i$ and then taking the metric
adjoint yield both mixed brackets in Eq.~\eqref{eq:signed-algebra}.
Positive signature recovers Eq.~\eqref{eq:witten-algebra}.

\smallskip
\noindent\appendixlabel{app:surface-composite}\textbf{The conserved composite and its polynomial.}
From here on impose $W=\Omega F$.
Let $E_{ij}$ be the constant matrix with a single unit in row $i$,
column $j$. The degree-zero to degree-two product and its reverse map are
\begin{equation}
 P=\tfrac12\Pi_2dc\Pi_0
   =\tfrac12(TX+\varepsilon Y^2)E_{41},\qquad
 P^\sharp=\tfrac\varepsilon2
          (X^\dagger T^\dagger+\varepsilon Y^2)E_{14}
          =\varepsilon P^\dagger.
 \label{eq:signed-composite-matrix}
\end{equation}
This also specifies the complete composite matrix: for
$\varepsilon=1$, $P=\mathcal Q^{(2)}_+$ of Eq.~\eqref{eq:descendant}.
The product is even under the original fermion parity and becomes odd
under the grading of the retained endpoint pair.

The first-order algebra determines its product polynomial without expanding
four derivatives. Since $d^\sharp$ and $c^\sharp$ annihilate zero-forms,
\[
 d^\sharp dc\Pi_0
 =2\mathcal H_\varepsilon c\Pi_0
      +2i\varepsilon\Omega Kd\Pi_0,\qquad
 c^\sharp d^\sharp dc\Pi_0
 =4\varepsilon\mathcal H_\varepsilon^2\Pi_0-4\Omega^2K^2\Pi_0.
\]

The corresponding two-form calculation gives the other product. On
degree zero and degree two, respectively,
\begin{equation}
 P^\sharp P=\varepsilon(\mathcal H_\varepsilon^{(0)})^2-\Omega^2K^2,
 \qquad PP^\sharp=\varepsilon(\mathcal H_\varepsilon^{(2)})^2-\Omega^2K^2.
\end{equation}
Since $P^\sharp=\varepsilon P^\dagger$, the positive-adjoint polynomial
on degree zero is $P^\dagger P=(\mathcal H_\varepsilon^{(0)})^2-\varepsilon\Omega^2K^2$.
Fixing $K=k$ therefore gives $C=-\varepsilon\Omega^2k^2$.
For negative signature this uses a new positive pairing on the retained
sectors, as explained in Section~\ref{sec:signature}; the change of
adjoint is not a unitary transformation of the indefinite parent.

\smallskip
\noindent\appendixlabel{app:middle-channels}\textbf{Resolving the scalar polarization channels.}
Set $Y=ik/F$ in Eq.~\eqref{eq:surface-middle-expanded}, conjugate by
$T_F=F^{1/2}$, and add $a$. With $g=\hbar F'/(2F)$,
Eq.~\eqref{eq:measure} gives the common diagonal correction
$\Delta_W-\hbar^2(F'/F)'=g^2-\hbar g'$. The off-diagonal part becomes
$-i\hbar kF'\mathcal R_\varepsilon/F^2$. These two observations give
exactly $H_{I,W}$ and $H_0$ in Eq.~\eqref{eq:surface-intermediate}.

For $k\ne0$, take $\beta^2=\varepsilon\Omega^2k^2$ and set
$r=-i\beta/(\Omega k)$, so $r^2=-\varepsilon$. The vectors and dual rows
\begin{equation}
 v_\beta=(r,1)^t,\qquad v_{-\beta}=(-r,1)^t,\qquad
 \ell_\beta=(1/(2r),1/2),\qquad
 \ell_{-\beta}=(-1/(2r),1/2)
\end{equation}
satisfy $\mathcal R_\varepsilon v_{\pm\beta}=\pm r v_{\pm\beta}$ and
$\ell_{\alpha}v_{\gamma}=\delta_{\alpha\gamma}$ for
$\alpha,\gamma\in\{\beta,-\beta\}$. Hence
$H_{I,W}v_\beta=v_\beta H_I(\beta)$ with
$H_I(\beta)=H_0-\hbar\beta W'/W^2$, reproducing the scalar
intermediate of Eq.~\eqref{eq:beta-chain}.

The first-order factors follow from the same projections. After removing
the radial measure, the raising blocks of $d_\chi$ are
\begin{equation}
 d_{01}=\begin{pmatrix}\hbar\partial_x+W-g\\ik/F\end{pmatrix},
 \qquad d_{12}=\begin{pmatrix}-ik/F&\hbar\partial_x+W+g\end{pmatrix}.
\end{equation}
Direct contraction gives
$\ell_\beta d_{01}=B_{1,\beta}/(2r)$ and
$d_{12}v_\beta=B_{2,\beta}$, with the factors of
Eq.~\eqref{eq:beta-factors}. Resolving $\mathcal R_\varepsilon$ into its
two eigenspaces in $d_{12}\mathcal R_\varepsilon d_{01}$ shows that
each channel contributes $B_{2,\beta}B_{1,\beta}/2=q_+$; together
they give $2q_+$. At $k=0$, the off-diagonal matrix term vanishes and
both scalar channels coincide, with $\beta=0$.

For $\varepsilon=1$ the polarization vectors are orthogonal in the
positive inner product. For $\varepsilon=-1$ they are real null vectors
of $G_1=\diag(1,-1)$, with $v_\beta^\dagger G_1v_{-\beta}=-2$.
The signed pairing therefore couples the two channels. Their generally
complex-conjugate potentials are consistent with metric self-adjointness
of the full matrix, but not with separate positive scalar adjoints.
For constant $W$ the term proportional to $W'$ vanishes: positive $C$
requires complex factorization energies, without necessarily requiring
a complex intermediate potential.

\smallskip
\noindent\appendixlabel{app:surface-global}\textbf{Global deformation on a torus.}
If $x$ as well as $\varphi$ is periodic and $F>0$, the condition
$W=\Omega F$ with $\Omega\ne0$ gives
$\oint W\dd x=\Omega\oint F\dd x\ne0$.
Thus the closed one-form $\alpha=W\dd x$ cannot equal $\dd\chi$
for a globally single-valued real $\chi$. Nevertheless,
$d_\alpha=\hbar\dd+\alpha\wedge$ is globally defined and nilpotent.
Its local expression is the same $d_\chi$ used above, so the matrix
formulas and their algebraic identities remain valid. Only the
representation by conjugation with $e^{\chi/\hbar}$ is local.
For ordinary periodic forms, this global distinction excludes zero
energy. In the $\ell=0$ sector, the formal zero modes in the order
$(0,1_x,1_\varphi,2)$ are proportional to
$e^{-\chi/\hbar}$, $F^{-1}e^{\chi/\hbar}$,
$F^{-1}e^{-\chi/\hbar}$ and $e^{\chi/\hbar}$.
None is periodic, since its multiplier around the $x$ cycle is
$\exp[\pm\Omega\oint F\dd x/\hbar]\ne1$.
For $\ell\ne0$, the central bound gives
$\mathcal H_W\geq|\Omega\hbar\ell|>0$.
The self-adjoint elliptic Hamiltonian on the compact torus therefore
has a strictly positive lowest eigenvalue in every form degree,
including the $C_0=0$ sector. This excludes zero modes, without
excluding saturation of a nonzero central bound.

\section{Pauli angular shifts and the conserved intermediate}
\label{app:pauli}

In the planar measure $r\dd r\dd\varphi$,
$\partial_r^\dagger=-\partial_r-1/r$ and
$Je^{i\varphi}=e^{i\varphi}(J+\hbar)$. The radial adjoint term and
angular shift cancel, giving
$R_+=e^{i\varphi}(-\hbar\partial_r+J/r+\omega r-\hbar\Gamma/(2r))$.
Composition with Eq.~\eqref{eq:pauli-first} yields
\begin{equation}
 [Q_\alpha,R_\beta]=0,\qquad [R_-,R_+]=4\hbar\omega,\qquad
 R_+R_-=2\mathcal H_P+4\omega(J-\hbar/2),\quad\alpha,\beta=\pm.
 \label{eq:pauli-check-algebra}
\end{equation}
Put $P_\pm=(1\pm\Gamma)/2$. Since
$Q_\pm Q_\mp=2\mathcal H_PP_\pm$, the products $S_+S_-$ and
$S_-S_+$ are $2\mathcal H_PP_\pm[2\mathcal H_P+4\omega(J\pm\hbar/2)]$,
respectively. On each selected spin sector, $J\pm\hbar/2=K$;
adding the products proves Eq.~\eqref{eq:pauli-composite}.

After the radial embedding specified in Section~\ref{sec:pauli},
the middle channel of Eq.~\eqref{eq:pauli-route} has Hamiltonian
\begin{equation}
 2h_I=-\hbar^2\partial_r^2+\omega^2r^2
       +\frac{k^2-\hbar^2/4}{r^2}-2\omega k+2\hbar\omega.
 \label{eq:pauli-intermediate}
\end{equation}
The endpoint blocks are $h_\pm=H_\pm-a-\omega k$, with $H_\pm$
in Eq.~\eqref{eq:pauli-radial-H}. Using
$(\hbar\partial_r+v)^\dagger(\hbar\partial_r+v)
=-\hbar^2\partial_r^2+v^2-\hbar v'$ for the real coefficients of
$T,V$ in Eq.~\eqref{eq:pauli-legs} gives
\begin{equation}
 \begin{aligned}
 T^\dagger T&=2h_-+4\omega k,&TT^\dagger&=2h_I+4\omega k,\\
 V^\dagger V&=2h_I,&VV^\dagger&=2h_+.
 \end{aligned}
 \label{eq:pauli-factor-check}
\end{equation}
Thus $h_IT=Th_-$ and $h_+V=Vh_I$, with the actual intermediate
angular channel retained. Expanding $VT$ gives Eq.~\eqref{eq:q}
with the data~\eqref{eq:pauli-target}, while
\begin{equation}
 q_-q_+=\tfrac14T^\dagger V^\dagger VT
       =\tfrac12T^\dagger h_IT
       =h_-(h_-+2\omega k)=(H_--a)^2-\omega^2k^2.
\end{equation}
Reversing the factors proves the upper-sector identity.
The charge product thus confirms $C=-\omega^2k^2$, previously
identified by comparing the reduced Hamiltonians with Eq.~\eqref{eq:H}.

For $k=\hbar\ell$, the coefficients of $r^{-2}$ in $2H_\pm$
are
\begin{equation}
 g_\pm=\hbar^2\big[(\ell\mp1)^2-\tfrac14\big],\qquad
 \nu_\pm=|\ell\mp1|.
 \label{eq:pauli-axis-indices}
\end{equation}
These are also the leading coefficients for the hyperbolic polar
reduction, because $F(r)=r+O(r^3)$ and $A_\varphi=O(r^2)$.
The endpoint criterion in Section~\ref{sec:cone} makes $H_+$
limit-circle only at $\ell=1$, and $H_-$ only at $\ell=-1$;
all other integer sectors are limit-point at the axis.
In a smooth, nonrotating spin frame near the axis the orbital labels
are $m_+=\ell-1$ and $m_-=\ell+1$. Regular surface spinors
select $u_\pm\sim r^{1/2+|m_\pm|}$, with the nonlogarithmic
$\sqrt r$ branch at $m_\pm=0$. Neither $g_+$ nor $g_-$ vanishes
for integer $\ell$. These are scalar endpoint conditions; common
domains for the unreduced charges require a separate analysis.

At $k=0$, both endpoints have $\nu=1$, and the planar Hamiltonians
coincide. They are limit-point at zero and infinity, so the minimal
operator on $C_c^\infty(0,\infty)$ has a unique self-adjoint closure.
Here $H_\pm-a=T^\dagger T/2$ and $q_+=q_-=-(H_\pm-a)$.
A singlet at $E=a$ would require $Tu=0$, but
$u_0\propto r^{-1/2}\exp[-\omega r^2/(2\hbar)]$ is not
square-integrable at the axis. For $\omega\ne0$, the levels are
$E_n-a=2\hbar|\omega|(n+1)$, $n=0,1,\ldots$;
for $\omega=0$, $a$ is a continuum threshold, not an eigenvalue.
The charge identity extends to the scalar closures; the corresponding
matrix charges remain odd.

\section{Checks for the hyperbolic Pauli reductions}
\label{app:hyperbolic}
\label{app:hyperbolic-polar}

\noindent\textbf{Polar reduction.}
For Eq.~\eqref{eq:hyperbolic-polar}, the identities
$F'^2-FF''=1$, $A_\varphi'=BF$ and $A_\varphi(1+F')=BF^2$
reduce the commutators $[R_-,Q_\pm]$ to zero; the composition must
include $Je^{-i\varphi}=e^{-i\varphi}(J-\hbar)$.
Hence $R_-$ also commutes with $\mathcal H_P$.
Removing the radial measure changes $\hbar\partial_r$ to
$\hbar\partial_r-\hbar F'/(2F)$. Along
Eq.~\eqref{eq:pauli-route} the two factors become
\begin{equation}
 T=\hbar\partial_r+\frac{k\cosh\rho+A_\varphi+\hbar/2}{F},
 \qquad V=\hbar\partial_r+\frac{k-\hbar/2-A_\varphi}{F},
 \qquad q_+=\tfrac12VT.
 \label{eq:hyperbolic-legs}
\end{equation}
To check their factorization energies, write
$a_j=\hbar\partial_r+(j-A_\varphi)/F$ and
$h_{-,j}=a_j^\dagger a_j/2$. The three channel Hamiltonians are
$h_-=h_{-,k+\hbar/2}$, $h_I=h_{-,k-\hbar/2}$ and
$h_+=a_{k-\hbar/2}a_{k-\hbar/2}^\dagger/2$.
Expansion using the identities above gives exactly
Eq.~\eqref{eq:pauli-factor-check} with $\omega k$ replaced by
$\beta_R=k(k+2BR^2)/(4R^2)$. In particular,
$T^\dagger T-2h_-=TT^\dagger-2h_I=4\beta_R$.
The same intertwining calculation as in Appendix~\ref{app:pauli}
therefore proves Eq.~\eqref{eq:hyperbolic-closure} and its upper-sector
counterpart.

The coefficient of $\hbar\partial_r$ in $VT$ is
$k(\cosh\rho+1)/F=2W$; its multiplication term gives
$U=\Delta_W+\beta_R^2/W^2$, establishing the data in
Eq.~\eqref{eq:hyperbolic-target}. At $k=0$, $V=-T^\dagger$ and
$\beta_R=0$, so the factorization identities give
$q_+=VT/2=-h_-=-h_+=-(H_\pm-a)$ for every uniform $B$.
These are equalities of radial differential expressions on $r>0$;
operator equalities require the corresponding domains. Here $W\equiv0$
makes the divided parametrization~\eqref{eq:correction} inapplicable,
but the factors remain well defined.
The spin structure extending through the axis gives
$k/\hbar\in\mathbb Z$. The endpoint classification and regularity
conditions are those of Eq.~\eqref{eq:pauli-axis-indices}; neither
inverse-square coefficient vanishes in an allowed angular sector.
Thus this polar reduction does not produce nonsingular full-line
partners by removing the endpoint.

In particular, at $k=0$ the scalar Hamiltonians have unique
self-adjoint closures, with $H_\pm-a=T^\dagger T/2\geq0$.
Since $T=\hbar\partial_r+\hbar/(2r)+O(r)$ near zero,
its formal zero mode behaves as $r^{-1/2}$ and is not normalizable.
At infinity the potential of $H_\pm-a$ tends to $B^2R^2/2$
with an exponentially decaying remainder. Its essential spectrum
is therefore $[B^2R^2/2,\infty)$, and for $B\ne0$ the lowest
spectral value lies strictly above zero. At $B=0$, zero is a
continuum threshold, not an eigenvalue. Thus $E=a$ is not a
physical singlet in this sector for any $B$.

The nonzero-$k$ cancellation $C=0$ requires
$B=-\hbar\ell/(2R^2)$. Its semiclassical behavior depends on what
is held fixed: at fixed nonzero $\ell$ and fixed $R$, the required
field tends to zero as $\hbar\to0$. Keeping $B\ne0$ and $R$ fixed
instead requires $\ell=-2BR^2/\hbar\in\mathbb Z$ along the
semiclassical sequence.

\smallskip
\noindent\appendixlabel{app:hyperbolic-geodesic}\textbf{The regular geodesic pair.}
For Eq.~\eqref{eq:hyperbolic-geodesic}, the unreduced charge is
$Q_+=\sigma_+[\hbar(\partial_x+F'/(2F))+(J-A_y)/F]$, where
$F=R\cosh(x/R)$ and $J=-i\hbar\partial_y$.
Removal of the measure gives Eq.~\eqref{eq:geodesic-A}.
At $B=-\hbar/R^2$ and $j=0$, put $w=\gamma\tanh(x/R)$,
$\gamma=\hbar/R$, $\epsilon=\gamma^2/2$ and $p=-i\hbar\partial_x$.
The identities $w^2+\hbar w'=\gamma^2$ and
$w^2-\hbar w'=\gamma^2[1-2\operatorname{sech}^2(x/R)]$ give the regular pair
\eqref{eq:regular-PT-pair}. Its normalized bound state is
$\phi_0=(2R)^{-1/2}\operatorname{sech}(x/R)$, with
$A_0\phi_0=0$ and $h_-\phi_0=0$.

Set $A=A_0$, $\mathcal H_0=\diag(h_+,h_-)$ and
$\Gamma=\diag(1,-1)$. The four Hermitian charges are
\begin{equation}
 \mathcal S_1=\begin{pmatrix}0&A\\A^\dagger&0\end{pmatrix},
 \quad \mathcal T_1=\begin{pmatrix}0&pA\\A^\dagger p&0\end{pmatrix},
 \qquad \mathcal S_2=i\Gamma\mathcal S_1,\quad
 \mathcal T_2=i\Gamma\mathcal T_1.
 \label{eq:exotic-charges}
\end{equation}
Their algebra follows from $A^\dagger A=2h_-$,
$AA^\dagger=2h_+$, $p^2=2(h_+-\epsilon)$ and $[p,h_+]=0$:
\begin{align}
 \{\mathcal S_a,\mathcal S_b\}&=4\delta_{ab}\mathcal H_0,
 &\{\mathcal T_a,\mathcal T_b\}
 &=8\delta_{ab}\mathcal H_0(\mathcal H_0-\epsilon),\nonumber\\
 \{\mathcal S_a,\mathcal T_b\}&=2\delta_{ab}\mathcal L,
 &\mathcal L&=\diag(2p h_+,A^\dagger pA),\quad a,b=1,2.
 \label{eq:exotic-algebra}
\end{align}
All four charges are odd and commute with $\mathcal H_0$;
$\mathcal L$ commutes with them as well. Its lower block is the
third-order Lax--Novikov integral, while its upper block is a
polynomial multiple of free momentum. Intertwining gives
$(A^\dagger pA)^2=8h_-^2(h_--\epsilon)$, and hence
\begin{equation}
 \mathcal L^2=8\mathcal H_0^2(\mathcal H_0-\epsilon).
 \label{eq:exotic-spectral}
\end{equation}
Each block is a Burchnall--Chaundy relation between commuting scalar
operators \cite{BurchnallChaundy1923}, distinct from the auxiliary
Bol equation of Section~\ref{sec:target}.

At the same field but general $j$,
\begin{equation}
 h_+(j)=\frac{p^2}{2}+\epsilon
 +\frac{j^2}{2R^2}\operatorname{sech}^2(x/R)
 +\frac{j\gamma}{2R}\operatorname{sech}(x/R)\tanh(x/R).
 \label{eq:geodesic-general-j}
\end{equation}
Although $A_j$ intertwines for every $j$, the defect of $pA_j$ is
$(pA_j)h_-(j)-h_+(j)pA_j=[p,h_+(j)]A_j$.
The nonconstant terms above vanish only at $j=0$, explaining why
this additional symmetry belongs to that Fourier sector.
This conclusion concerns the specific product $pA_j$, without
excluding another possible unrestricted lift.

\section{Verification of the positive matrix chain}
\label{app:matrix}

Put $L=\hbar\partial_x+W$. For the maps in Eq.~\eqref{eq:ST},
the first ODE~\eqref{eq:matrix-ode} gives
\begin{equation}
 TS=(L-z)(L+z)+(\lambda-\eta)(\lambda+\eta)
 =L^2+\hbar z'-z^2+\lambda^2-\eta^2=L^2-U=2q_+.
\end{equation}
The second ODE is $e'=0$, with $e$ defined in
Eq.~\eqref{eq:matrix-chain-constant}.
In $H_IS-SH_-$ and $H_+T-TH_I$, all coefficients of derivatives
of the wavefunction cancel. The first components vanish using $e'=0$
and the derivative of the first ODE. The remaining component of each
defect is $\hbar E$, where
\begin{align}
 E&=-\frac\hbar2\eta''+
       \left(z-\frac{W\eta}{\lambda}\right)\eta'
       +\frac{W'}\lambda(\lambda^2-\eta^2),\nonumber\\
 \lambda E&=\frac{\hbar^2}{4}W'''-UW'-\frac12WU'
           =\frac{\hbar^2}{4}\Bol[W].
 \label{eq:matrix-E}
\end{align}
The expression for $\lambda E$ follows by differentiating the two ODEs
and eliminating $z'',\eta''$. Both intertwinings therefore hold whenever
the target satisfies $\Bol[W]=0$.

Expanding $S^\dagger S$ and $TT^\dagger$, then eliminating $z'$
with the first ODE and $Wz+\lambda\eta$ with $e$, gives
\begin{equation}
 S^\dagger S=2(H_--a+e+\lambda^2),\qquad
 TT^\dagger=2(H_+-a-e+\lambda^2).
 \label{eq:matrix-norms}
\end{equation}
If the local chain extends to closed operators with these identities,
positivity requires $H_-\geq a-e-\lambda^2$ and
$H_+\geq a+e-\lambda^2$. Thus $e,\lambda$ and the boundary
conditions must be compatible with the endpoint spectra; local ODE
existence alone does not establish such a global realization.
By contrast, the second-derivative coefficient of $SS^\dagger$ and
$T^\dagger T$ is $-\hbar^2\diag(1,0)$, rather than
$-\hbar^2\id_2$ as in $2H_I$. Neither product can therefore be
replaced by $2H_I$ plus a scalar constant. This is the precise
obstruction to applying the scalar factorization argument to this chain.

\section{The constant positive-invariant example}
\label{app:constant}

For constant $W=w\ne0$ and $U=-\mu^2$, the one-dimensional
charge factors as $2q_+=B_2B_1$ with
$B_{1,2}=\hbar\partial_x+w\pm i\mu$.
Both factors commute with the common Hamiltonian, but their adjoint
norms contain opposite terms linear in momentum. They therefore do
not satisfy the scalar Schr\"odinger factorization assumptions of
Eq.~\eqref{eq:real-chain}.

Use the operators of Eqs.~\eqref{eq:constant-D}--\eqref{eq:constant-G}
with the standard positive adjoint and fermionic relations
$\{\epsilon_i,\epsilon_j\}=0$,
$\{\epsilon_i,\epsilon_j^\dagger\}=\delta_{ij}$.
Since $X,Y$ commute with each other and their adjoints, the fermionic
relations give nilpotency and the brackets stated in the main text;
in particular,
$\{D_1,D_2^\dagger\}=-XY^\dagger+YX^\dagger
=2\hbar(\mu\partial_x+w\partial_y)=Z$.
Set $Q_1=D_1$, $Q_2=iD_2$,
$\widehat H=\mathcal G-a+\mu^2$ and $Z_H=-iZ$.
The four Hermitian first-order generators and their brackets are
\begin{gather}
 R_i=\frac{Q_i+Q_i^\dagger}{\sqrt2},\qquad
 S_i=\frac{Q_i-Q_i^\dagger}{i\sqrt2},\qquad i=1,2,\nonumber\\
 \{R_i,R_j\}=\{S_i,S_j\}
 =2\delta_{ij}\widehat H+(1-\delta_{ij})Z_H,\qquad
 \{R_i,S_j\}=0.
 \label{eq:constant-algebra}
\end{gather}
They commute with both $\mathcal G$ and the Hermitian translation
$Z_H$, giving centrally extended linear $N=4$ supersymmetry.
With $P_x=-i\hbar\partial_x$ and $P_y=-i\hbar\partial_y$, the
central combinations are explicitly positive:
\begin{equation}
 \widehat H\pm\frac{Z_H}{2}
 =\frac12\big[(P_x\pm\mu)^2+(P_y\pm w)^2\big]\geq0.
 \label{eq:constant-central-positivity}
\end{equation}
They are the squares of $(R_1\pm R_2)/\sqrt2$ and, likewise,
of $(S_1\pm S_2)/\sqrt2$.

On $M=\mathbb R_x\times S^1_{L_y}$, Schwartz functions in $x$
with finite Fourier sums in $y$, tensored with $\mathbb C^4$, form a
common invariant core. Fourier transformation turns each $R_i,S_i$
into a Hermitian matrix multiplier. Since $R_i^2=S_i^2=\widehat H$,
their self-adjoint closures have Sobolev domain $H^1(M)\otimes\mathbb C^4$;
$\mathcal G$ has domain $H^2(M)\otimes\mathbb C^4$.
The core is dense in these graph norms and is preserved by all
charges and their products. The algebra therefore has a global
positive operator realization, with product identities on compatible
domains, without requiring each charge to preserve all of $D(\mathcal G)$.
For periodic forms, $P_y$ has eigenvalues $2\pi\hbar n/L_y$,
$n\in\mathbb Z$. Consequently both central combinations in
Eq.~\eqref{eq:constant-central-positivity} have purely absolutely
continuous spectrum $[\delta^2/2,\infty)$, where
$\delta=\operatorname{dist}(w,(2\pi\hbar/L_y)\mathbb Z)$.
Even when $\delta=0$, the zero-energy threshold has no normalizable
state, because the longitudinal momentum remains continuous.

The polar unitary of Eq.~\eqref{eq:polar-unitary} is particularly
explicit here. In longitudinal momentum space,
$H_\pm(p)-a=(p^2+w^2-\mu^2)/2$ and
$q_+(p)=[(w+ip)^2+\mu^2]/2$, so
\begin{equation}
 V(p)=\frac{(w+ip)^2+\mu^2}
 {\sqrt{(p^2+w^2-\mu^2)^2+4w^2\mu^2}},\qquad |V(p)|=1.
 \label{eq:constant-polar-unitary}
\end{equation}
Its denominator never vanishes for $C>0$. This unitary commutes
with the common Hamiltonian and preserves its domain, but is not a
finite-order differential operator. It is the polar factor of $q_+$;
the identical endpoint Hamiltonians also admit the trivial unitary
identification.

\smallskip
\noindent\textbf{Allowing a variable radius.}
For the extension considered in Eq.~\eqref{eq:constant-extension-defect},
use the orthonormal coframe $(\dd x,f\dd y)$ and the measure
$f\dd x\dd y$. Set $b=\hbar f'/f$,
$X=\hbar\partial_x+W$, $T=X+b$ and
$Y=f^{-1}(\hbar\partial_y-\mu)$. The successive blocks of
$d_\alpha$ are $(X,Y)^t$ and $(-Y,T)$; those of
$d_\alpha^c$ are $(-Y,X)^t$ and $(-T,-Y)$.
The identity $TY=YX$ gives $d_\alpha^2=(d_\alpha^c)^2=0$;
the displayed blocks also give $\{d_\alpha,d_\alpha^c\}=0$.
The adjoints are
$X^\dagger=-\hbar\partial_x-b+W$,
$T^\dagger=-\hbar\partial_x+W$ and
$Y^\dagger=f^{-1}(-\hbar\partial_y-\mu)$.
The zero- and two-form Hamiltonian blocks coincide. The difference
on one-forms has diagonal entry
$(T^\dagger T-XX^\dagger)/2=-\hbar s_f$ and off-diagonal entry
\begin{equation}
 \frac12\big(T^\dagger Y-YX^\dagger+Y^\dagger T-XY^\dagger\big)
 =b(Y+Y^\dagger)=-\frac{2\hbar\mu f'}{f^2}.
\end{equation}
The other diagonal entry is $\hbar s_f$, proving
Eq.~\eqref{eq:constant-extension-defect}. The resulting restriction
to constant $f,W$ applies to this pair of differentials, not to all
possible positive first-order constructions or to the local matrix
chains of Section~\ref{sec:limits}.

\end{document}